\documentclass[11pt,prd,letterpaper,showpacs,groupedaddress,superscriptaddress,nofootinbib,floatfix,preprintnumbers,tightenlines]{revtex4}
\usepackage{amssymb,amsmath,graphicx,xcolor,cancel}
\usepackage{bm,comment,slashed}

\begin{document}

\title{Machine learning action parameters in lattice quantum chromodynamics}

\author{Phiala E. Shanahan}
\affiliation{Department of Physics, College of William and Mary, Williamsburg, VA 23187-8795, USA}
\affiliation{Jefferson Laboratory, 12000 Jefferson Avenue, Newport News, VA 23606, USA}

\author{Daniel Trewartha}
\affiliation{Jefferson Laboratory, 12000 Jefferson Avenue, Newport News, VA 23606, USA}

\author{William Detmold}
\affiliation{Center for Theoretical Physics, Massachusetts Institute of Technology, Cambridge, MA 02139, USA}

\date{\today}

\preprint{MIT-CTP/4980}


\begin{abstract}
Numerical lattice quantum chromodynamics studies of the strong interaction are important in many aspects of particle and nuclear physics. Such studies require significant computing resources to undertake.
A number of proposed methods promise improved efficiency of lattice calculations, and access to regions of parameter space that are currently computationally intractable, via multi-scale action-matching approaches that necessitate parametric regression of generated lattice datasets.
The applicability of machine learning to this regression task is investigated, with deep neural networks found to provide an efficient solution even in cases where approaches such as principal component analysis fail. The high information content and complex symmetries inherent in lattice QCD datasets require custom neural network layers to be introduced and present opportunities for further development.
\end{abstract}

\maketitle

\section{Introduction}

Lattice quantum chromodynamics (LQCD)~\cite{Wilson:1974sk} is a well established numerical method~\cite{Rothe:1992nt,Gattringer:2010zz} used to study quantum chromodynamics (QCD), the theory of the strong interaction. A central part of the Standard Model (SM) of nuclear and particle physics, strong interactions bind quarks and gluons into protons and nuclei, and dictate the emergence of complex nuclear structure in nature.
High-precision LQCD calculations are important in determining the parameters of the SM and guide searches for evidence of new physics beyond it~\cite{Aoki:2016frl}. Recent LQCD calculations also provide new insights into the quark and gluon structure of protons~\cite{Constantinou:2015agp} and the structure and interactions of light nuclei~\cite{Beane:2010em,Davoudi:2017ddj}. Similarly, LQCD calculations have enabled investigations of QCD matter at extreme temperatures, and efforts to understand QCD matter at high density are underway~\cite{Ding:2015ona}.
These calculations are extremely computationally demanding, consuming significant fractions of the computational resources that are available for scientific research worldwide.

LQCD calculations are performed on a discrete 4-dimensional space-time grid (typically a hypercubic lattice), and use Monte-Carlo importance sampling~\cite{Luscher:2010ae} to determine the dynamics of the quark and gluon fields defined on this space. Achieving physical results requires a series of calculations at different discretisation scales (referred to as the lattice spacing), and different lattice volumes, and a subsequent extrapolation to the continuum (where the discretisation vanishes) and infinite volume limits. 
Particularly challenging is the approach to the continuum limit; the computational cost of the Hybrid Monte-Carlo (HMC) algorithm \cite{Duane:1987de} typically used scales with a high inverse power of the lattice spacing, $a$, approximately $a^{-z}$ with $z> 6$ for a fixed physical lattice volume~\cite{Schaefer:2010hu}. Known as critical slowing down, this occurs because of the quasi-local nature of the HMC updating procedure, requiring an increasing number of steps to update physics on a fixed physical volume as the lattice spacing decreases. A number of methods attempt to circumvent this issue by acting at multiple physical length scales. Examples include perfect actions~\cite{Bell:1975wtp,Hasenfratz:1993sp,Wiese:1993cb,Hasenfratz:1998gu} that aim to achieve almost-continuum physics at finite lattice spacings, and multi-scale thermalisation techniques \cite{PhysRevLett.56.1015,EDWARDS1991289,EDWARDS1992621,PhysRevD.50.6998,Endres:2015yca,Detmold:2016rnh}. Such approaches require careful renormalisation group matching \cite{Wilson:1973jj,Balaban:1989ra} of the LQCD actions defined at different scales such that they describe the same long-distance physics. An essential challenge is to solve the parametric regression task: Which action parameters best represent the coarse-scale physics of an ensemble of samples generated at a finer resolution, and vice-versa? Similar parameter regression problems of LQCD datasets arise in the context of mixed action LQCD simulations (see for example Ref.~\cite{Schroers:2003mf,Bowler:2004hs,Beane:2005rj}).

In this work, machine learning (ML) techniques, in particular neural networks, are applied to the regression problem of determining LQCD action parameters from an ensemble of samples. 
Significant progress in ML over the last few years has led to new scientific applications of ML tools, including to a number of statistical and quantum mechanics problems. In one set of studies, ML has been used to infer the presence of phase transitions and thermodynamic properties in simple condensed matter models  \cite{2016arXiv160807848B,2017arXiv170302369L,2016arXiv160902552C,2016PhRvB..94p5134T}. In another study, variational methods have been optimized for many-body problems using ML techniques \cite{2017Sci,2016PhRvB..94s5105W}. Novel approaches to the Monte-Carlo method  that is ubiquitous in numerical simulations of many systems have also been developed using ML ideas \cite{2017PhRvB..95c5105H,2017PhRvB..95d1101L,2017NatPh..13..431C,2017PhRvB..96r4410W,2017arXiv170809401Z,2017arXiv170208586W,Tanaka:2017niz}.
Finally, ML regression for matching Hamiltonians in condensed matter contexts has recently been investigated~\cite{2017PhRvB..95d1101L,2018arXiv180101127S} and shows promise.
Very few studies, however, have applied ML techniques to investigate gauge field theories such as LQCD (LQCD is a particularly important example of a more general class of theories defined with a local invariance known as a gauge symmetry), and new techniques and adaptations are required because of the unique and complex symmetry structures of these theories\footnote{Ref.~\cite{Wetzel:2017ooo} investigates the ability for neural networks to learn a simple order parameter in pure SU(2) gauge theory at finite temperature.}. 
Averaged over Monte-Carlo importance sampling, LQCD data is invariant under discrete spacetime translations and hypercubic group transformations, although individual samples do not have these symmetries.
In addition, internal symmetries based on the continuous Lie group SU(3) associated with each spacetime location must be respected.
Exploiting these symmetries is essential to the success of the approach used here; it is found that suitably customised deep neutral networks can provide an efficient and practical method of determining the action parameters describing the physics of a given set of configurations. 

This article is arranged as follows. In Section \ref{sec:lqcd}, the basic aspects of the lattice QCD calculations that are used to train and test parametric regression by neural networks are discussed, and a principal component analysis (PCA) is used to ascertain the difficulty of the regression tasks that are attempted. In Section \ref{sec:ML}, a number of different neural network structures are studied. First, in Section \ref{sec:MLlinks}, a fully connected neural network is used. This easily solves the parameter regression problem on training ensembles, but suffers from over-fitting due to the inverted hierarchy of the information content of each sample to the number of samples available for training. Despite its failure to generalise, this network finds features that persist in the LQCD data for Monte-Carlo times considerably longer than those seen for typical physics-motivated observables. The over-fitting problem is remedied in Section \ref{sec:MLloops}, where several custom symmetry-enforcing layers are introduced to define neural network structures that efficiently solve the regression problem. The trained networks correctly resolve parameter differences even between ensembles which are essentially indistinguishable under the PCA analysis. 
Section \ref{sec:summary} provides a summary.
Two appendices provide additional details of aspects of machine learning and of the lattice QCD calculations.

\section{Lattice QCD}
\label{sec:lqcd}
Lattice QCD calculations are performed by approximating the QCD path integral by a Monte Carlo sum over gauge field configurations on a discrete four-dimensional space-time. The expectation value of an operator $\mathcal{O}$ that defines some physical quantity is given by:
\begin{eqnarray}
\label{eq:pathintegral}
\langle\mathcal{O}\rangle &=& \frac{1}{\mathcal{Z}}\int \mathcal{D}\psi \mathcal{D}\bar{\psi} \mathcal{D} A\, \mathcal{O}[\psi,\bar{\psi},A]\,e^{-S[\psi,\bar{\psi},A]} \\
&=& \frac{1}{\mathcal{Z}}\int \mathcal{D} U \tilde{\mathcal{O}}[U]\,e^{- \tilde S[U]},
\end{eqnarray}
where $\mathcal{Z} = \int \mathcal{D}\psi\mathcal{D} \bar{\psi} \mathcal{D} A\, e^{-S[\psi,\bar{\psi},A]}$, the (anti-)fermion and gluon fields (gauge fields) are denoted by $\psi(\bar{\psi})$ and $A$, and $S[\psi,\bar{\psi},A]$ is the discretised QCD action (defined in Appendix~\ref{app:lattaction}). In the second line, the fermion and anti-fermion fields are integrated out exactly, and the gauge fields are transformed to link fields $U=e^{i A}$, to give an effective action $\tilde S[U]$ and operator $\tilde{\mathcal{O}}[U]$ depending only on the gluon link fields.
The resulting integral can be approximated as 
\begin{eqnarray}
\label{eq:pathintegral}
\langle\mathcal{O}\rangle
\approxeq \frac{1}{N_\text{cfg}} \sum_{i=1}^{N_\text{cfg}} \mathcal{O}[U_i],
\end{eqnarray}
where the gauge field configurations $U_i$ ($i$ indexes the configurations in a given ``ensemble'' of fields) are distributed according to the probability measure $e^{-\tilde S[U]}$. In practice, this is guaranteed by sampling the fields from a Markov chain Monte-Carlo stream for which this probability measure is a fixed point. These representative gauge fields are the input data for the ML approaches to parametric regression studied here. For additional details of the LQCD approach, see Refs.~\cite{Rothe:1992nt,Gattringer:2010zz} and Appendix~\ref{app:lattaction}.

Lattice QCD gauge fields are represented as links between sites on a 4-dimensional lattice of volume\footnote{The spatial, $L$, and temporal, $T$, extents of the lattice geometry are often distinct.} $V = L^{3} \times T$, with the lattice sites separated by some physical distance $a$, typically 0.05--0.15 fm. Each link, labelled by $U_{\mu}(x)$, where $x$ denotes the spacetime coordinates of the origin site and $\mu$ the direction of the link, is encoded by an SU(3) matrix (a $3\times3$ complex matrix $M$ with $M^{-1}=M^\dagger$ and $\det[M]=1$)\footnote{Here, $M^\dagger = (M^{\ast})^{T}$ is the Hermitian conjugate. An SU(3) matrix can be specified by 8 real numbers, but typically the redundant representation with 18 real numbers is used.}. Links in opposing directions are related via $U_{-\mu}(x) = U^{\dagger}_{\mu}(x-\hat{\mu})$, and only links in the positive direction are stored. In this format, a gauge field used in typical modern lattice QCD calculations, where for example $L=64$ and $T=128$, is described by $L^3\times T \times 4 \times 18 \approx \mathcal{O}(10^9)$ floating point or double precision numbers, where the factor of 4 arises from the number of positive spacetime directions (labelled by $\mu$). In order to recover QCD results, calculations must be performed on a number of ensembles of field configurations with different lattice spacings $a$ and lattice volumes $V$, and the continuum ($a\to0$) and large-volume ($V\to\infty$) limits must be taken.

The governing equations of QCD and their lattice counterparts have a variety of symmetries, some that are highly non-trivial. The symmetries satisfied by ensembles of gauge fields are of particular interest in the context of the ML approaches studied here, as they place strong restrictions on numerical operations that can be performed on lattice data to extract physically meaningful results.
In particular, lattice QCD is invariant under a local symmetry of the gauge fields known as a gauge symmetry; this is an invariance under local multiplications of link variables by SU(3) matrices
\begin{equation}
U_{\mu}(x) \rightarrow U'_{\mu}(x) = \Omega(x)U_{\mu}(x)\Omega^{\dagger}(x + \hat{\mu}) \,\,\, \mathrm{  for \,\,all } \,\,\, \Omega(x) \in \text{SU(3)},
\end{equation}
referred to as a gauge transformation (note that the matrix $\Omega(x)$ differs at every spacetime point). This symmetry is not apparent from the numerical representation of a QCD configuration, but rather constrains physical observables calculated on a given gauge field to be invariant under all gauge transformations of that field.
In addition, lattice QCD defined on a discretised finite volume is invariant under discrete translations and under 4-dimensional rotations and reflections (transformations generated by the hypercubic group, $H_4$ \cite{Mandula:1983ut}). 
Unlike gauge symmetry, these latter symmetries do not hold on a configuration-by-configuration basis, but rather
emerge after averaging physical quantities over all gauge fields in an ensemble.
An additional important property of QCD is that a characteristic length scale, $1/\Lambda_{\rm QCD}\sim 1$ fm, emerges dynamically from the interactions of the theory, setting a spacetime distance over which values of the link fields are correlated. 

\subsection{Lattice QCD ensembles}
\label{sec:ensembles}

A number of different ensembles of lattice QCD gauge field configurations were used for this first exploratory study. Each ensemble was generated using a two-colour $N_c=2$ Wilson gauge action with $N_f=2$  flavours of dynamical Wilson fermions (defined in Appendix~\ref{app:lattaction}). This action depends on two bare couplings/parameters, $\beta$ and $m_0$. QCD with $N_c=2$ exhibits similar rich dynamical structure to the full theory with $N_c=3$ and is a natural testing ground for the new approaches developed here. Ensembles were generated with a standard HMC algorithm using a leapfrog integrator to take molecular dynamics trajectory steps of length $\tau_{MD}=0.5$ in 15--40 substeps (tuned to keep the acceptance  rate $\sim70$\%). In each case, the streams were initialised from a hot start or from a thermalised lattice from a nearby set of couplings, and the initial 500 trajectories were not included in the further analysis. For most ensembles, configurations were saved every 10 trajectories to generate ensembles of $\mathcal{O} (10^{3})$ independent configurations, with the separation determined from studies of the autocorrelation times of typical observables (for some ensembles, configurations were saved every trajectory to allow studies of autocorrelation times to be undertaken). 
Since $N_c=2$ in these calculations, rather than $N_c=3$ in  full QCD, the lattice data structures used here are somewhat smaller than those used for state-of-the-art calculations, with each configuration represented by ${\cal O}( 10^{6})$ double precision numbers. 
All ensembles were generated using a modified version of the {\tt chroma} lattice field theory library \cite{Edwards:2004sx} that was previously~\cite{Detmold:2014kba} found to produce results consistent with an independent code \cite{Lewis:2011zb}. 

\begin{figure}
	\includegraphics[width=0.65\columnwidth]{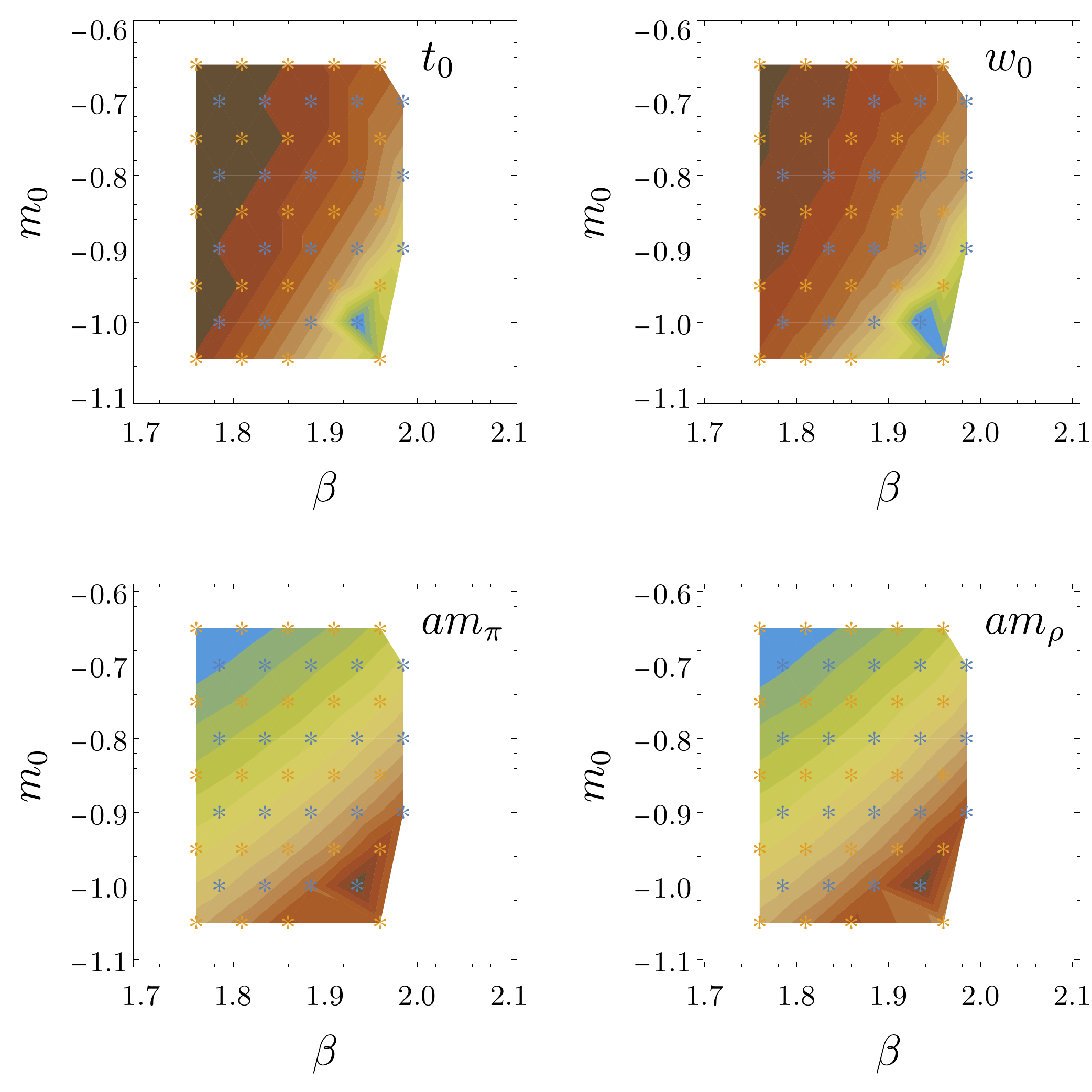}
	\caption{\label{fig:physcontourplot} 
		Contours show the scale setting quantities $t_0$ and $\omega_0$, as well as the lattice spacing times the pion mass $am_\pi$, and rho meson mass $am_\rho$, determined using calculations on each ensemble in the two $L/a=12$ grids. The stars show the locations of the ensembles from Grids A (blue) and B (orange).
	}
\end{figure}

Ensembles were generated at many points in parameter space:
\begin{itemize}
	\item Grid A: Twenty $12^3\times 36$ ensembles of 10,000 trajectories with each $\beta\in\{1.785,1.835,1.885,1.935,1.985\}$ and 
	$m_0\in\{-0.7,-0.8,-0.9,-1.0\}$, excluding the pair $\{\beta,m_0\}=\{1.985,-1.0\}$ which could not be thermalised efficiently; 
	\item Grid B: Twenty-five $12^3\times 36$ ensembles of 10,000 trajectories with each $\beta\in\{1.76,1.81,1.86,1.91,1.96\}$ and $m_0\in\{-0.65,-0.75,-0.85,-0.95,-1.05\}$, excluding the pair $\{\beta,m_0\}=\{1.91,-1.05\}$ which could not be thermalised efficiently; 
	\item Grid C: Twenty ensembles with the same bare parameters as Grid A, but with a spacetime volume of $16^3\times 48$, excluding the pairs $\{\beta,m_0\}=\{1.935,-1.0\}$ and $\{1.985,-1.0\}$, which could not be thermalised efficiently;
\item
Two sequences of ensembles with parameters tuned to produce closely matched plaquette values. The parameters of each set are indicated by the parentheses $(\beta,m_0)$:
	\begin{itemize}
		\item Set D: $\{ D_1(1.815,-0.98),\, D_2(1.825,-0.93),\, D_3(1.838,-0.87), \,D_4(1.85,-0.83)$\\
		 $D_5(1.862,-0.79)\}$;
		\item Set E: $\{E_1(1.826,-1.03),\,E_2(1.837,-0.99),\, E_3(1.847,-0.95),\,E_4(1.858,-0.9)$ \\
		$E_5(1.87,-0.85)\}$;
	\end{itemize} 
\item Set F: Ten independent streams of 10,000 trajectories denoted $F_1,\ldots,F_{10}$, saved every trajectory, generated with the same values of $\beta=1.76$ and $m_0=-0.75$. 
\end{itemize}
Simple physical observables, including the pion and rho meson masses and scale setting observables $w_0$ and $t_0$ \cite{Luscher:2010iy}, have been calculated on Grids A and B; contour plots displaying the variation of these quantities across the ensembles are shown in Fig.~\ref{fig:physcontourplot}.

In order to check the validity of the HMC streams, the evolution of simple quantities along the trajectories has been monitored.
The simplest, and computationally cheapest, way to produce a gauge invariant quantity from links is to take the trace of products of links over closed loops (``Wilson loops''). Wilson loops are defined from gauge links as shown schematically in Fig.~\ref{fig:loopfig}, and detailed in Appendix~\ref{app:lattaction}. Planar Wilson loops $W_{k\times l}(x)$, with indices $k$ and $l$ denoting the dimensions of the loop (with orientation label suppressed), were computed for square loops up to $6 \times 6$, as well as rectangular loops of size $1\times n$ for $n=2,\ldots,12$, and all possible planar orientations. The evolution of representative loop types for the ensembles in Grids A, B, and C, averaged over orientations and spacetime position, is shown in Appendix~\ref{app:ensemblevalid}. For each case, this evolution indicates that the data is well thermalised after approximately 500 trajectories.

\begin{figure}
	\includegraphics[width=0.4\textwidth]{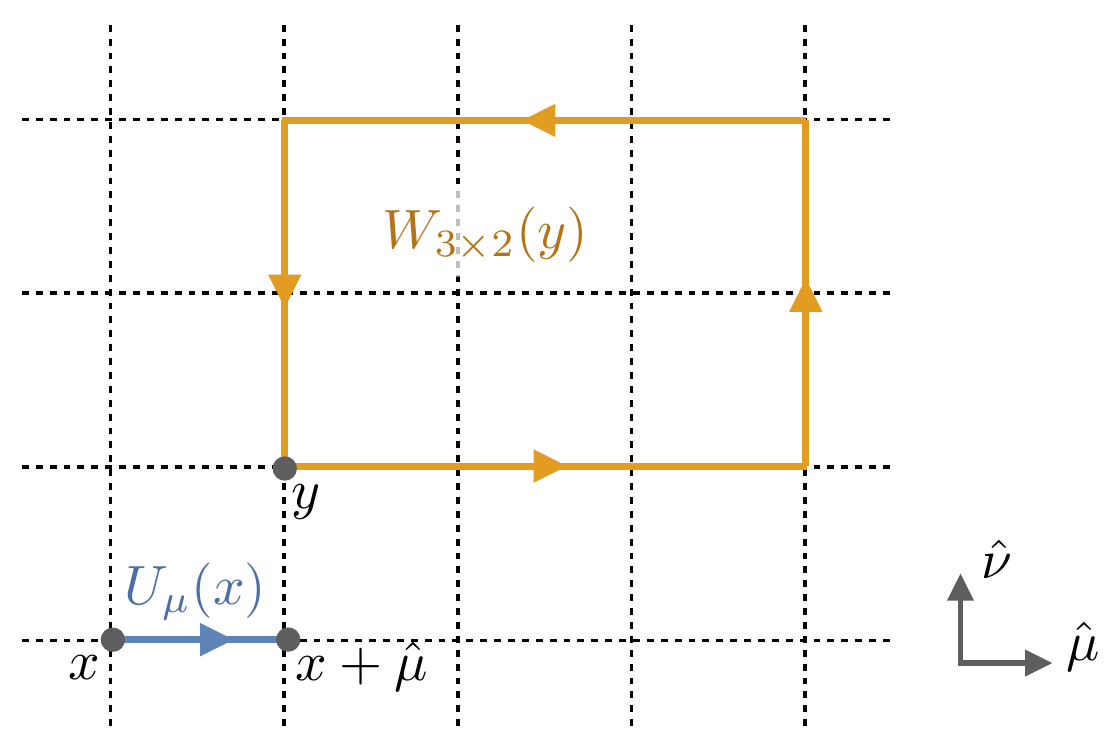}
	\caption{\label{fig:loopfig}Diagrammatic representation of the construction of planar Wilson loops $W_{k\times l}(x)$, with indices $k$ and $l$ denoting the dimensions of the loop (with orientation label suppressed), from gauge links $U_{\mu}(x)$. }
\end{figure}

To determine the number of HMC steps required for gauge field configurations to be independent, the autocorrelation times of  the pion and rho two-point correlation functions, and of the same sets of Wilson loops introduced above, have been calculated. 
The autocorrelation function for a given operator $\mathcal{O}$ is defined as 
\begin{equation}\label{eq:rho}
\rho(\tau) = \sum_{\tau^\prime}\langle ( \mathcal{O}(\tau^\prime)-\langle\mathcal{O}\rangle)(\mathcal{O}(\tau^\prime+\tau)-\langle \mathcal{O} \rangle ) \rangle,
\end{equation}
where $\tau$ is the trajectory difference in the autocorrelation. This function  decays exponentially as $\rho(\tau) \sim \text{exp}[-\tau/\tau_\text{exp}]$ at large Monte-Carlo times $\tau$.
The decay constant $\tau_\text{exp}$ defines an autocorrelation time. Calculations of the autocorrelation time using this definition can suffer from large uncertainties, especially when $\tau_\text{exp}$ is small.
Another definition of the autocorrelation time is \cite{Madras:1988ei,Gattringer:2010zz} 
\begin{equation}\label{eq:tau}
\tau_\text{int}=\frac{1}{2} + \lim_{\tau_{\rm max}\to\infty}\frac{1}{\rho(0)}\sum_{\tau=0}^{\tau_{\rm max}} \rho(\tau),
\end{equation}
 which approaches a constant as $\tau_{\rm max}\to\infty$.
The autocorrelation functions and integrated autocorrelation times $\tau_\text{int}$ for the Wilson loops, and those for the zero-momentum projected pion and rho two point correlation functions, $C_{\pi(\rho)}$ (defined in Appendix \ref{app:lattaction}), are shown in Fig.~\ref{fig:corrtimepirho}. 
In all cases, the integrated autocorrelation time is $\lessapprox 10$ trajectories, validating the choice to take trajectories spaced by this distance as an uncorrelated set to form an ensemble. Other observables may have different autocorrelation times, but the observables considered here are relatively representative\footnote{The topological charge of the gauge field typically has a long autocorrelation time, but at the relatively coarse lattice spacings used here, it will be comparable to that of the observables that are investigated.}.

\begin{figure}
	\centering
	\includegraphics[width=1.3\textwidth]{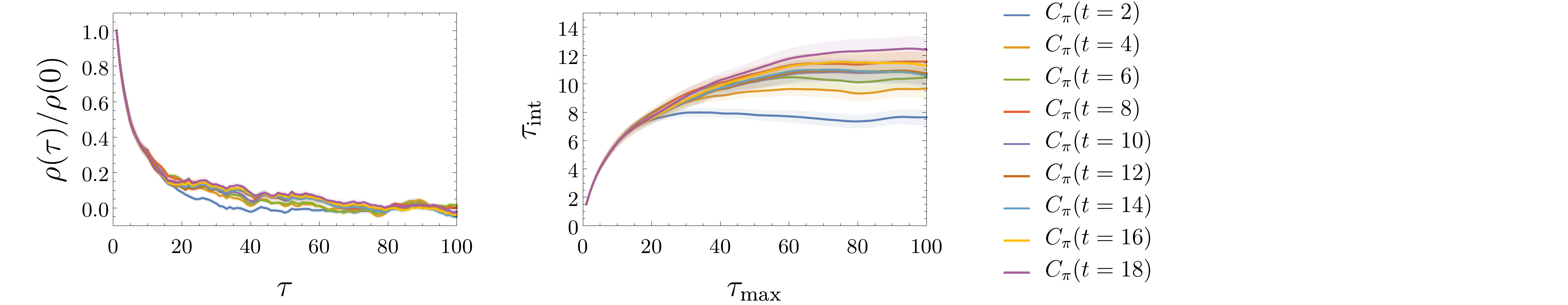}
 \\
 {~}
 \\
  \includegraphics[width=1.3\textwidth]{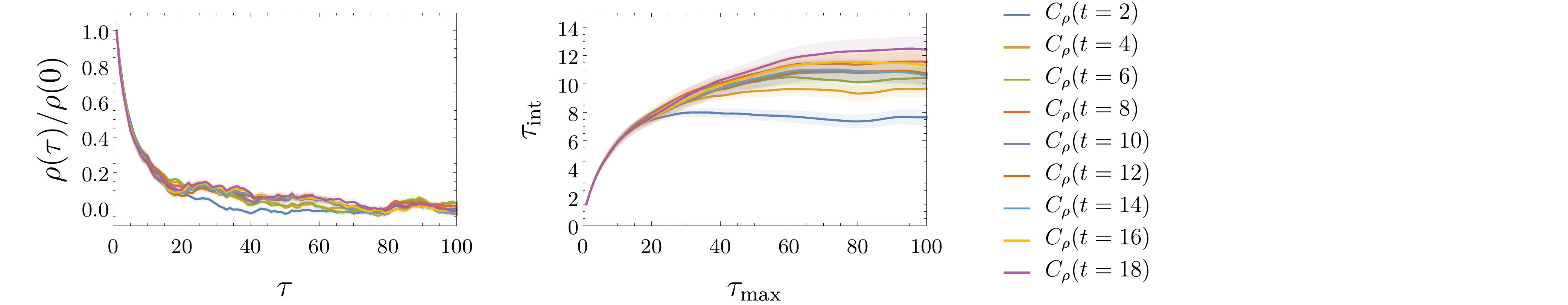}\\
  	\includegraphics[width=1.3\textwidth]{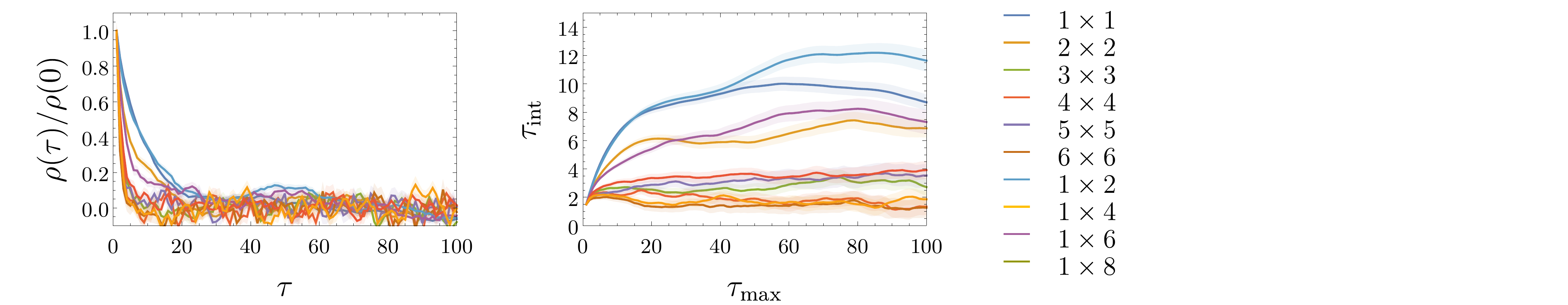}
	\caption{\label{fig:corrtimepirho}Autocorrelation functions $\rho(\tau)/\rho(0)$ (left, defined in Eq.~\eqref{eq:rho}) and autocorrelation times $\tau_\text{int}$ (right, defined in Eq.~\eqref{eq:tau}) of the pion (top) and $\rho$ (centre) two-point correlation functions at different Euclidean time separations, and of the various space-time averaged $n\times m$ planar Wilson loops (bottom). Measurements are performed on a subset of ensemble F1, for $N_{\rm traj}=4000$ sequential trajectories ($N_{\rm traj}=7980$ for the loops). The colours identify the type of loop and the shaded bands correspond to the uncertainties on these quantities as determined from a bootstrap procedure using $N_{\rm boot}=100$ bootstrap resamplings of size $N_{\rm traj}$.}
\end{figure}

\subsection{Ensemble discrimination using principle component analysis}
\label{sec:PCA}

\begin{figure}
	\includegraphics[width=\columnwidth]{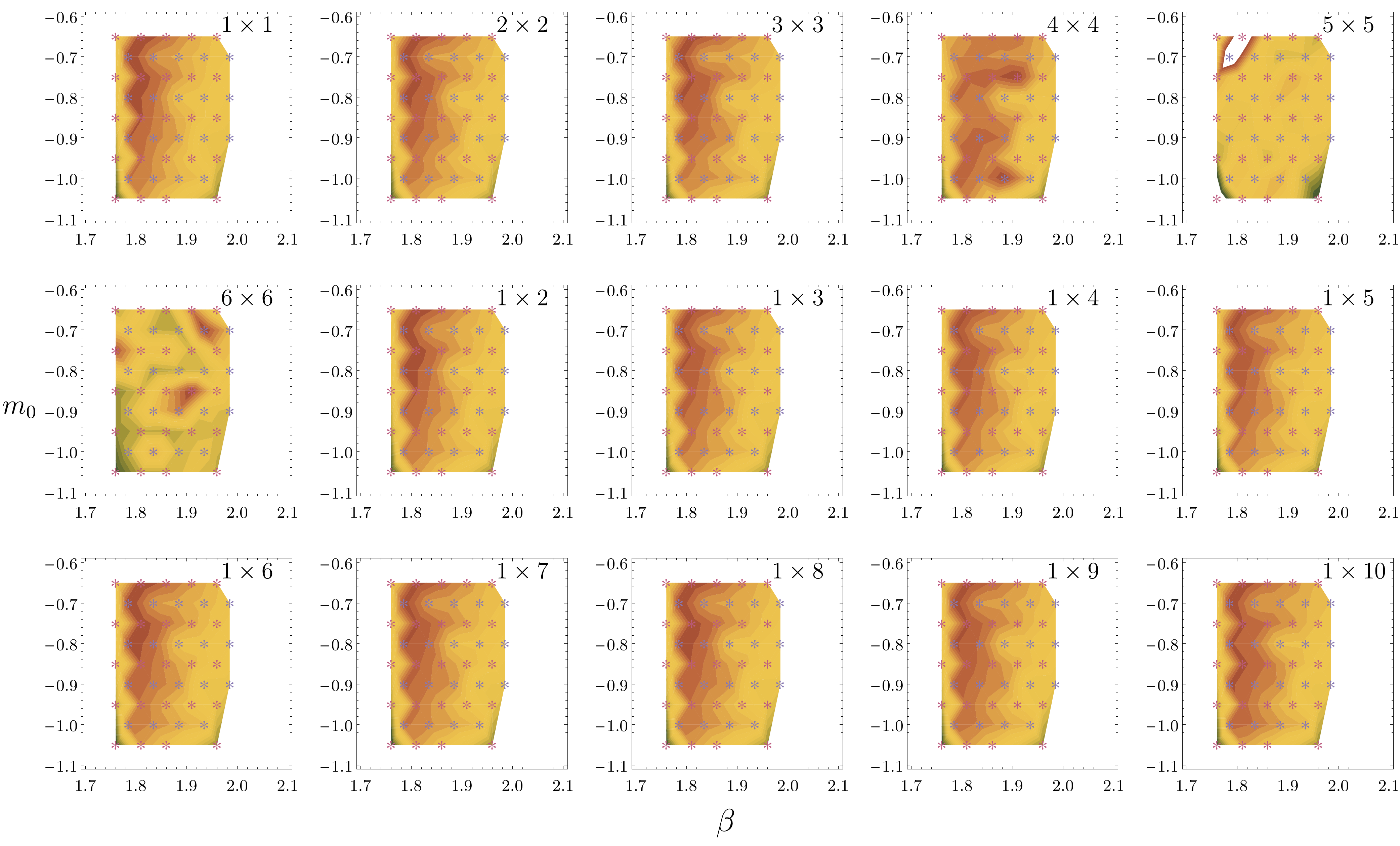}
	\caption{\label{fig:loopscontourplot} 
		Contours show $\ln|W_{n\times m}|$ from evaluations on each ensemble in the two $L/a=12$ grids. The stars show the locations of the ensembles from Grids A (blue) and B (orange).
	}
\end{figure}

To guide the application of ML methods to parametric regression of gauge fields in the space defined by the sample ensembles, the differentiability of the ensembles was assessed using a principle component analysis (PCA) \cite{doi:10.1080/14786440109462720,Hotelling:1933AA,Hotelling:1936AA}. Since Wilson loops  are  the simplest gauge-invariant objects, 
the  basis for the PCA was generated by calculating a set of square planar loops of sizes up to $L/2 \times L/2$, as well as $1\times n$ for $n$ up to $L$, averaged over all possible planar orientations and  space-time locations. Averaged loops are denoted
$W_{j\times l} = \sum_{\mathcal{O}(j\times l)} \sum_{x} W_{j\times l} (x)$, where the sum over $\mathcal{O}(j\times l)$ is over all hypercubic transformations of the indicated loop.
The averaged loop data are sufficiently small in dimension that it is possible to display them for a representative set of ensembles. Fig.~\ref{fig:loopscontourplot} shows contour plots of $\ln|W_{n\times m}|$ from evaluations on each ensemble in the two $L/a=12$ grids (Grids A and B). Figs.~\ref{fig:loophistsA}, \ref{fig:loophistsB}, and \ref{fig:loophistsC} (in Appendix~\ref{app:ensemblevalid}) show histograms for a subset of the loops for each ensemble in each of Grid A, B, and C, respectively. Clearly, some of the loops are statistically well determined, and subsets of the ensembles can be clearly distinguished. Ensembles in Grid C have loop distributions that are more sharply defined than those in Grids A and B as their larger spacetime volume enables more statistical averaging. For large loop sizes, all ensembles become hard to distinguish.

\begin{figure}
	\centering
	\includegraphics[width=0.8\textwidth]{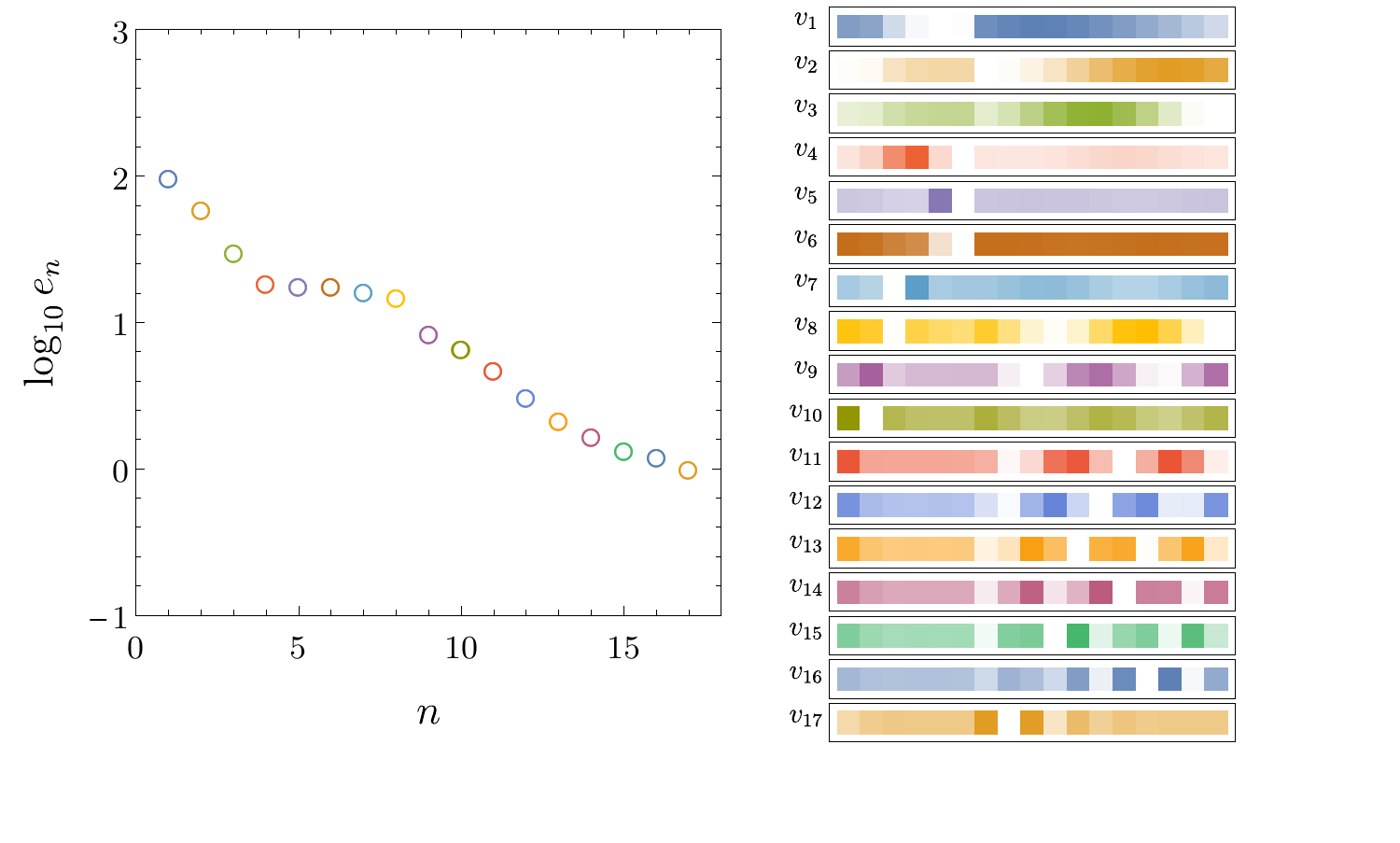}
	\caption{\label{fig:corrmat}Eigenvalues $e_n$ (left panel) and eigenvectors $v_n$ (right panel) of the loop correlation matrix for Grid A. The strength of the contribution of each loop to each eigenvector is represented by the tone of the corresponding box in the right panel (i.e., darker = larger contribution).}
\end{figure}

\begin{figure}
	\centering
	\includegraphics[width=0.8\textwidth]{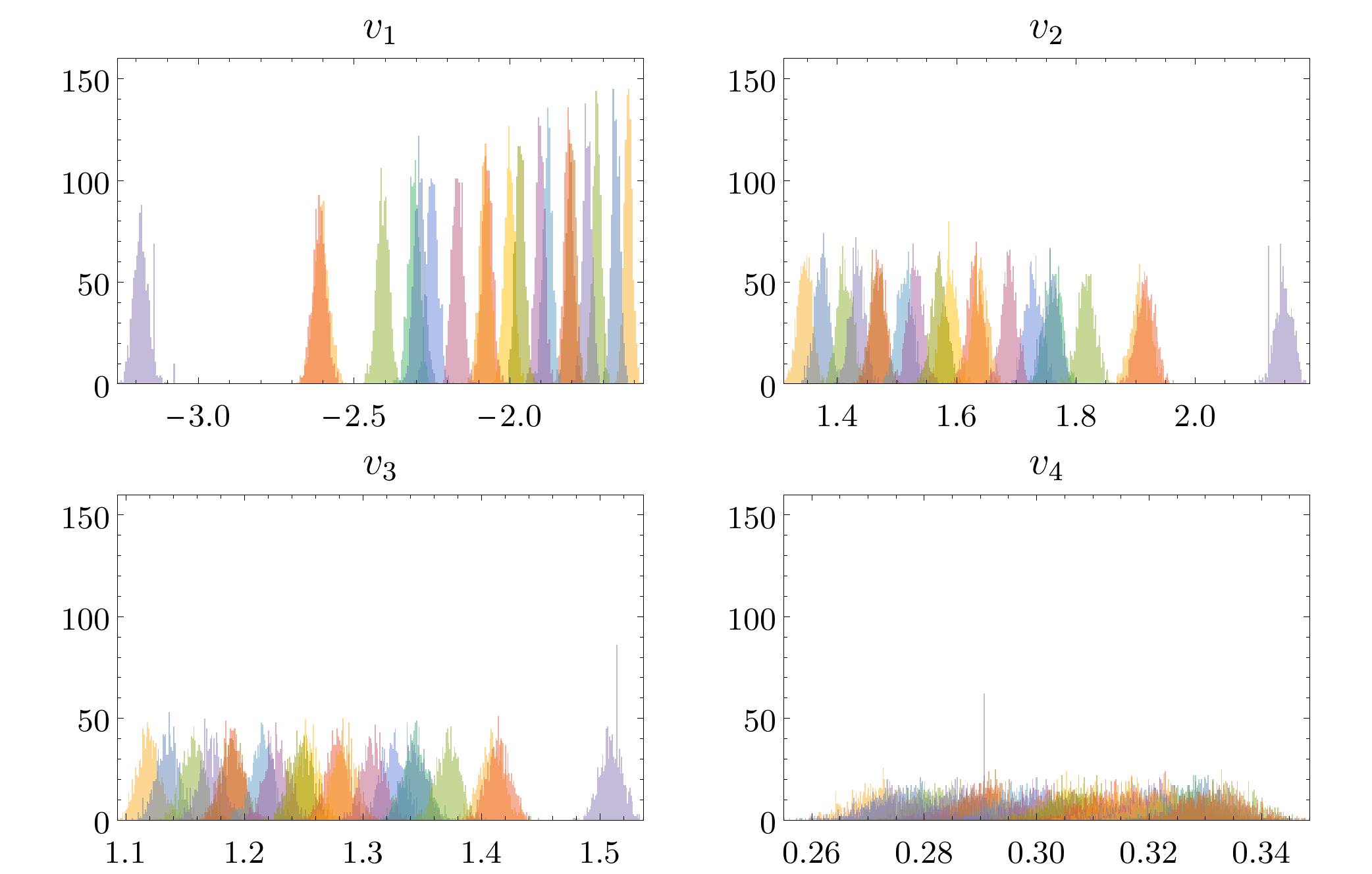}
	\caption{\label{fig:evechists}Combinations of loops corresponding to the four largest eigenvectors of the loop correlation matrix for Grid A. Each colour denotes a different ensemble in Grid A.}
\end{figure}

To perform the PCA on the loop data, a correlation matrix between the various loop observables can be constructed, either for a given ensemble, or, as is done here, across a collection of ensembles. The correlation matrix elements are
\begin{equation}
{\cal M}_{\ell_i,\ell_j}=\sum_e\sum_c \frac{\left[W_{\ell_i }(e,c) - \overline W_{\ell_i}(e)\right]\left[W_{\ell_j }(e,c) - \overline W_{\ell_j}(e)\right]}{\sigma(W_{\ell_i}(e))\sigma(W_{\ell_j}(e))}, 
\end{equation}
where $\ell_i \in \{1\times 1,2\times 2,\ldots\}$,
and $e$ and $c$ label the ensemble and the configuration in that ensemble, respectively. The summation over ensembles is for all ensembles in a given grid, and $\overline{X}$ and $\sigma(X)$ denote the mean and standard deviation of the given quantity over the particular ensemble of configurations.
The eigenvalues, $e_i$, and eigenvectors, $v_i$, of this correlation matrix for Grid A are shown in Fig.~\ref{fig:corrmat}. There are three particularly large eigenvalues. Similar pictures emerge from PCAs run on Grid B and Grid C, indicating three dominant degrees of freedom in the calculated Wilson loops.
Histograms showing the combinations of loops corresponding to the three dominant, and fourth sub-dominant, eigenvectors are presented  for the ensembles in Grid A in Fig.~\ref{fig:evechists}. 
Clearly, the information encoded in a collection of the simplest gauge-invariant objects is sufficient to distinguish all but a few of the ensembles in Grid A. 

The Jensen-Shannon divergence~\cite{Lin:1990AA,Lin:1991AA} provides a measure of the overlap of probability distributions and can be used to quantify the distinguishability of such distributions. 
Given two probability distributions $P$ and $Q$, defined over a space $X$, the Jensen-Shannon divergence is given by
\begin{equation}
D_{JS}(P||Q) = \frac{1}{2}D_{KL}(P||M)+\frac{1}{2}D_{KL}(Q||M),
\end{equation}
where $M = \frac{1}{2}(P+Q)$, and $D_{KL}(P||Q)$ is the Kullback-Leibler divergence \cite{kullback1951}, defined as
\begin{equation}
D_{KL}(P||Q) = \int dx \, P(x) \log_2 \frac{P(x)}{Q(x)}.
\end{equation}
The Jensen-Shannon divergence is bounded by $0\le D_{JS}(P||Q) \le 1$,
with $D_{JS}=0$ if and only if $Q=P$ almost everywhere, and larger values denoting lower overlap between distributions. The square root of the Jensen-Shannon divergence provides a well-defined metric \cite{1207388,Osterreicher:2003AA}.

\begin{figure}
	\centering
	\includegraphics[width=0.55\textwidth]{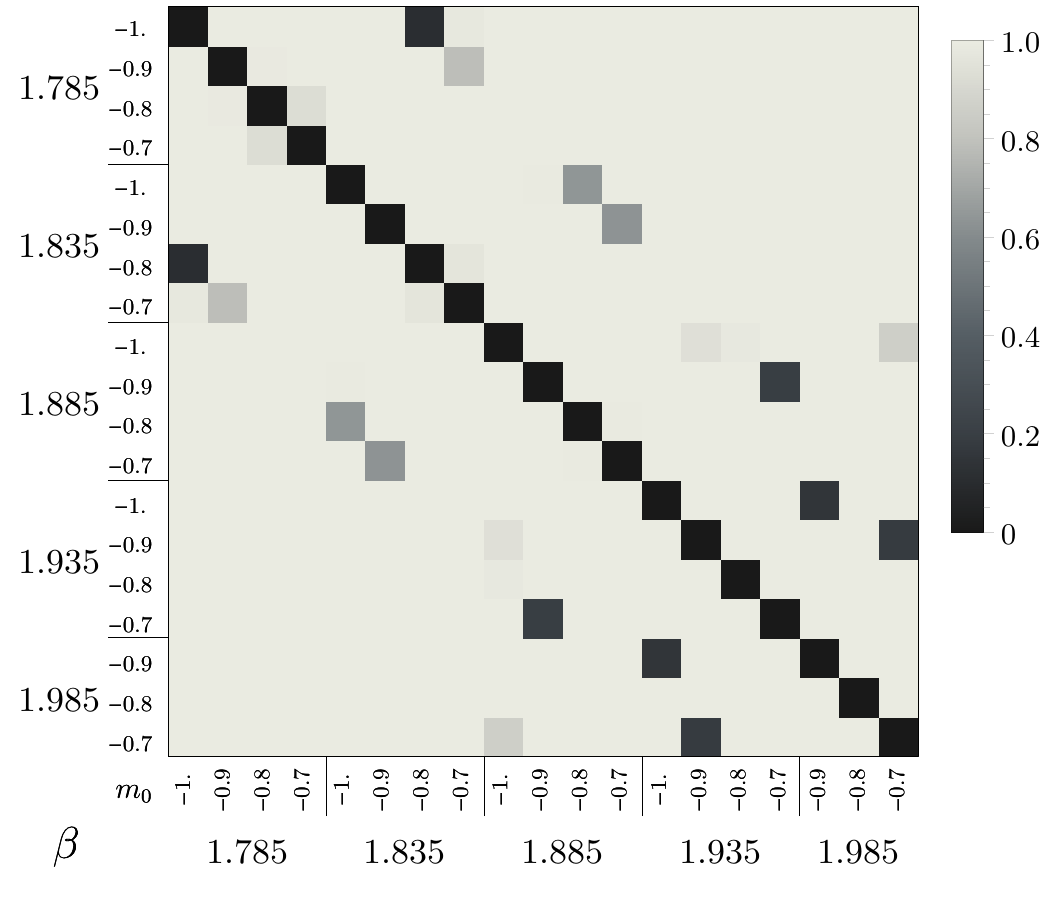}
	\caption{\label{fig:JS_PCA}The Jensen-Shannon divergence, $D_{JS}$,  between pairs of ensembles in Grid A, calculated over the three-dimensional distributions defined by the three dominant eigenvectors of the loop correlation matrix used for the PCA. $D_{JS}=1$ implies completely distinguishable distributions.}
\end{figure}

As a test of differentiability, the Jensen-Shannon divergences were calculated between all pairs of three-dimensional probability distributions defined by the three dominant eigenvectors of the loop correlation matrix for each ensemble in Grid A\footnote{On a given ensemble $e$, this three-dimensional probability distribution is given by
$P_e(s_i,s_2,s_3)$ 
where
$$s_i = v_i \cdot \left( W_{1\times1}(e,c), W_{2\times2}(e,c),  \ldots W_{1\times12}(e,c) \right),$$
and where $v_i$ is the $i$th eigenvector of the PCA.
Additional tests with the largest two or four eigenvectors gave qualitatively similar results.
}. 
To do this, each distribution was first interpolated over the samples from the given ensemble using smooth kernel distributions. The resulting values of $D_{JS}$ are shown pictorially in Fig.~\ref{fig:JS_PCA} for all pairs of the 19 ensembles in Grid A. Clearly, the dominant eigenvectors in loop space allow excellent differentiation between most pairs of ensembles, with approximately 8 out of 171 independent pairs that are only weakly, or not at all, differentiable. 

\begin{figure}
		\centering
		\includegraphics[width=0.8\textwidth]{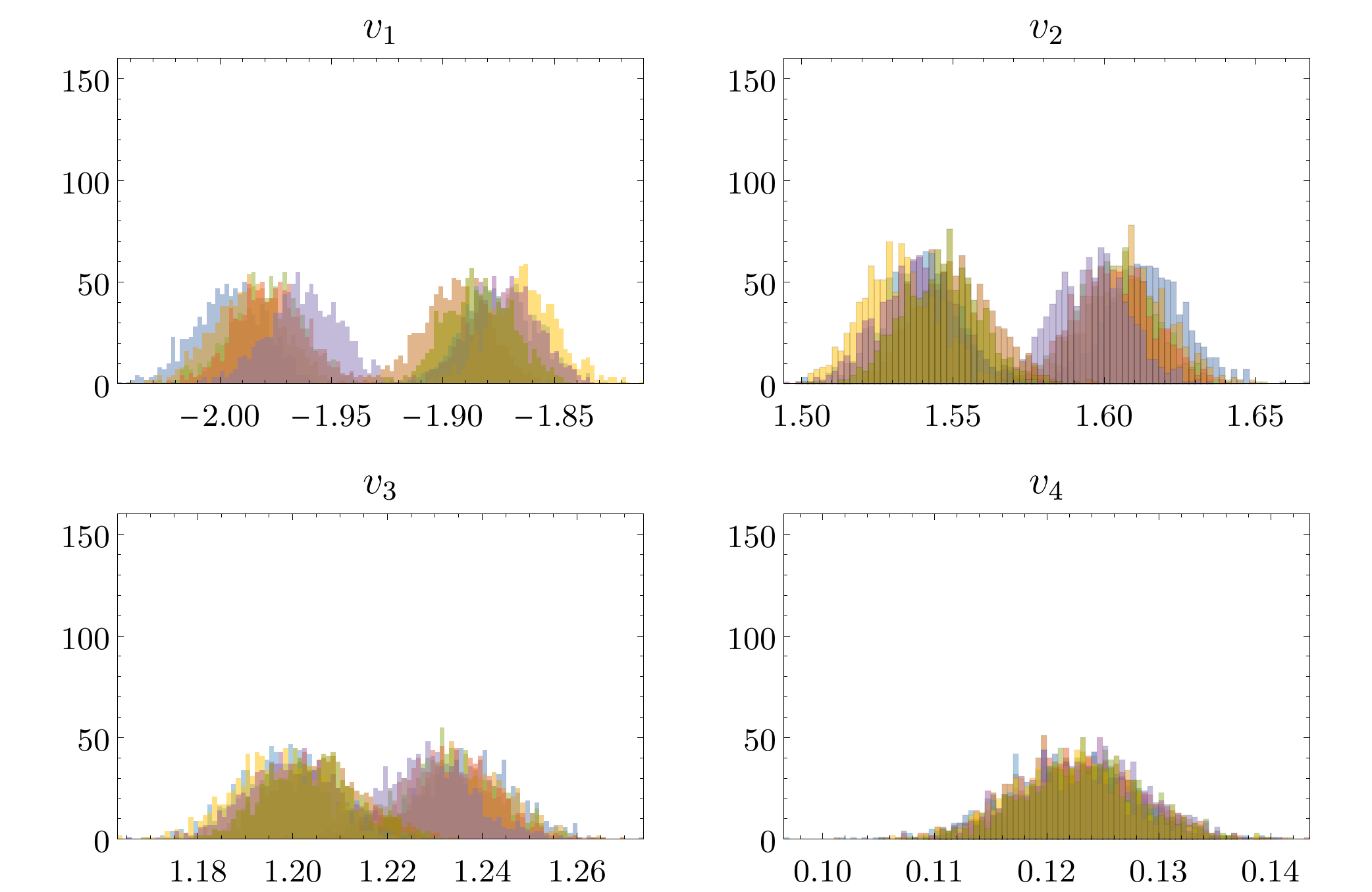}
		\caption{\label{fig:isopcahistograms}{Combinations of loops corresponding to the dominant eigenvectors of the loop correlation matrix for ensemble Sets D and E. Each colour denotes a different ensemble.}}
\end{figure}

\begin{figure}
	\centering
	\includegraphics[width=0.65\textwidth]{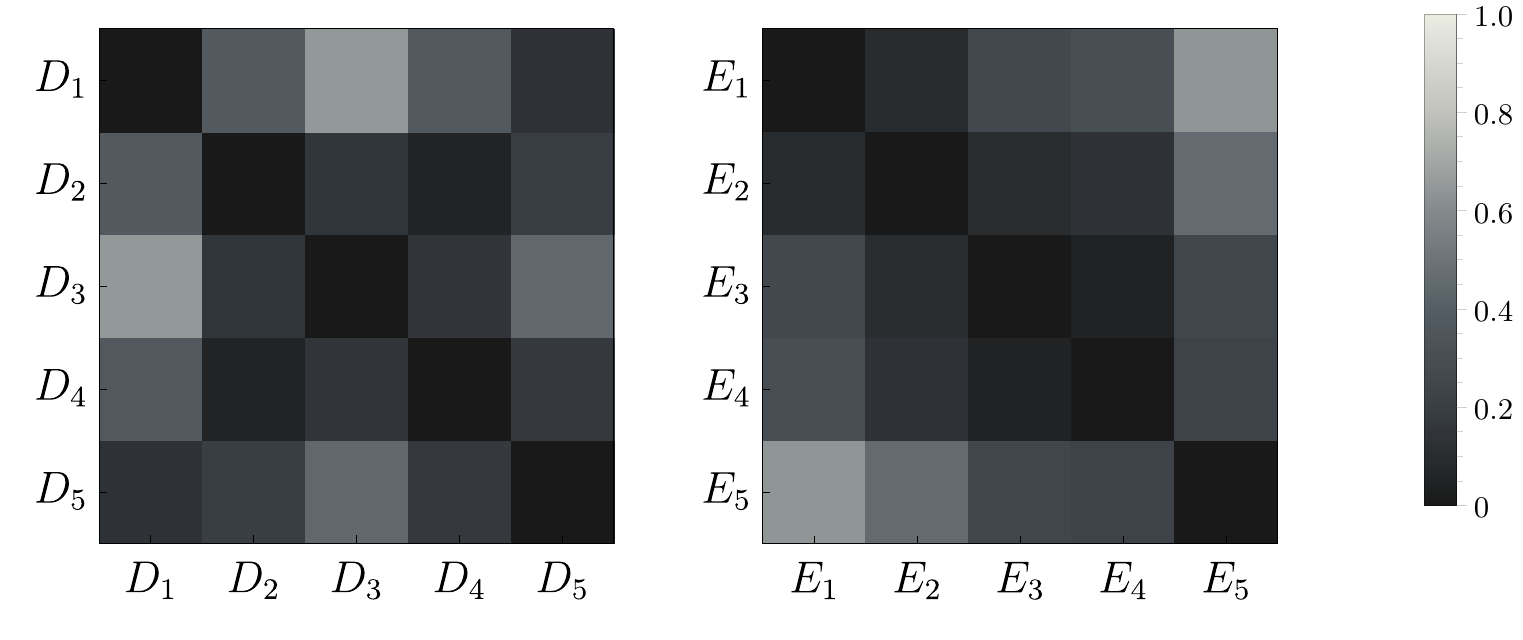}
	\caption{\label{fig:JS_PCA3}The Jensen-Shannon divergence, $D_{JS}$, between pairs of ensembles in Sets D (left) and E (right), calculated over the three-dimensional distributions defined by the three dominant eigenvectors of the loop correlation matrix used for the PCA. The maximum value of $D_{JS}$ in each Set is 0.6. $D_{JS}=1$ corresponds to completely distinguishable distributions.}
\end{figure}

A more challenging test of distribution differentiability is provided by the ensembles in Sets D and E, each designed to have maximal overlap of Wilson loops on each of the ensembles in the set, but different parameters in the $\{\beta,m_0\}$ plane. Fig.~\ref{fig:isopcahistograms} shows histograms of the combinations of Wilson loops corresponding to the dominant eigenvectors of the loop correlation matrix for ensemble Sets D and E, while Fig.~\ref{fig:JS_PCA3} displays the Jensen-Shannon divergence between pairs of ensembles in these sets. As the ensembles in each of Sets D and E are very poorly distinguishable in the space of Wilson loops, accurate differentiation between them presents a key challenge to parametric regression via ML.

\section{Neural networks for parametric regression of lattice QCD gauge fields}

\label{sec:ML}

Machine learning techniques, and in particular neural networks, offer a promising solution to parameter regression problems. The main focus of this work is to address such a problem in the context of LQCD: given an ensemble of lattice gauge fields, determine the parameters of a given action that are most likely to have generated it. As discussed in the introduction, this challenge arises, for example, in attempts to ameliorate critical slowing down by the matching of coarse and fine lattice actions, and in the context of perfect actions. Its solution will allow for more efficient LQCD calculations, enabling studies in regions of parameter space which are currently computationally unreachable.

To determine the action parameters of a given ensemble (for a particular choice of lattice action), one possible approach is to calculate a sufficiently large set of physics observables both on that ensemble and on a set of ensembles for which the parameters are known, and perform an interpolation and matching task using the calculated observables.
The alternative considered here is to train a neural network to perform the regression directly.
In principle, this approach is far more general than one based on a set of physics quantities, as the network can use all of the information encoded in a gauge field configuration. On the other hand, this is also challenging. As discussed in Section~\ref{sec:ensembles}, a single gauge field configuration is represented by $\mathcal{O}(10^9)$ real numbers in modern lattice QCD calculations. 
In comparison, a typical ensemble used for such calculations consists of $\mathcal{O}(10^3)$ configurations. This hierarchy implies that the stochastic learning of features of the relevant degrees of freedom of the gauge field configurations---in particular that extracted physics results must be invariant under spacetime translations, reflections, and hypercubic rotations as well as under gauge transformations---is challenging.

This challenge is approached in two ways, described in the following two sections. First, a multi-layer perceptron (a fully connected feed-forward neural network) is trained to learn the action parameters corresponding to lattice gauge field configurations. As anticipated, using gauge fields as input with no symmetry constraints leads to over-fitting of the spacetime and gauge features of the data which are not related to the physics encoded by a given ensemble. Nevertheless, this exploration reveals a number of interesting features of the problem at hand. Second, a practical solution to the parametric regression problem is provided in the form of a network with a structure that imposes the spacetime and gauge symmetries of LQCD (or, equivalently, involves preprocessing gauge field data into a format that respects these symmetries).

\subsection{Fully-connected network}

\label{sec:MLlinks}

The simplest approach to the parametric regression of lattice QCD gauge fields using neural networks is to use a multi-layer perceptron  \cite{Rosenblatt:1961AA,Werbos:1974AA,Linnainmaa:1970AA,Rumelhart:1986AA}, i.e., a fully-connected feed-forward network structure (a glossary of neural network terminology is provided in Appendix~\ref{app:glossary}). 
For each of the ensembles of gauge-field configurations in Grid B, 850 configurations were randomly selected as training data, while 100 were held out as validation data \cite{Mosteller:1968AA,Stone:1974AA}. Each gauge field configuration, consisting of $\mathcal{O}(10^6)$ real numbers, was treated as an individual input.
As physical quantities are only defined on ensemble average, regression on these inputs can not be exact; a given gauge configuration could, with various probabilities, have been generated from an action differing in both form and parameters from the one that it was in fact generated with, so a perfectly functioning network will necessarily have some spread in predictions from a given ensemble. 
Quantifying this maximum resolution is possible in principle, but computationally prohibitive, and for this reason has not been undertaken. Investigations into new ensemble-based training approaches that would sharpen the maximum regressor predictability are ongoing.

\begin{figure}
	\centering
	\includegraphics[width=0.8\textwidth]{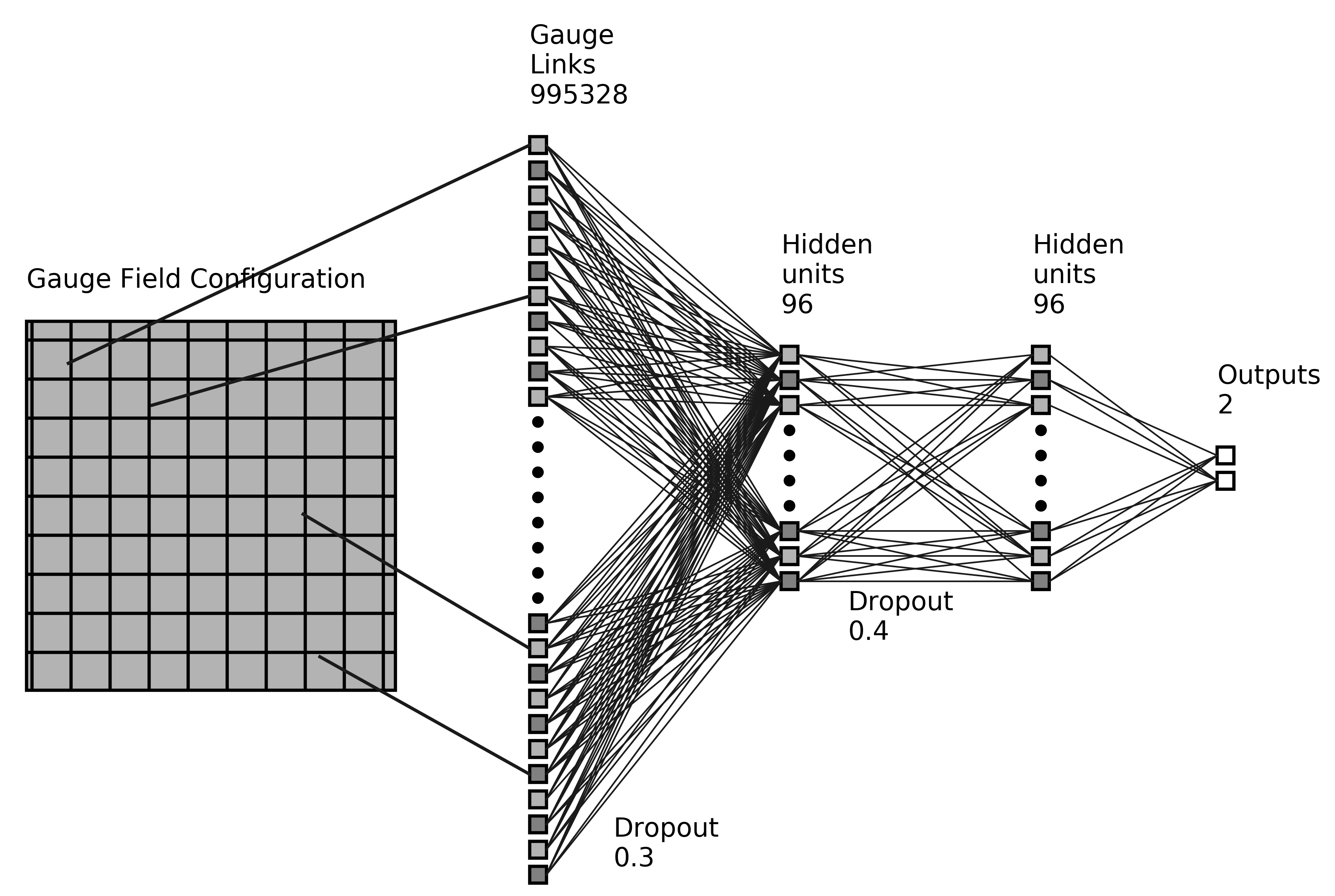}
	\caption{\label{fig:fcnn}A schematic representation of the neural network structure used for parametric regression. Gauge links, expressed in an SU(2) basis as 4 real numbers, are used as inputs to the network. There are 4 links in each positive direction from a given site, giving a total of $4 \times 4 \times V = 16 \times 12^{3} \times 36 = 995328$ real numbers per gauge field. Two fully connected layers, each with 96 nodes, are used. Each hidden layer features a {\tt tanh} activation function and dropout. A random set of connections between layers are omitted to denote dropout.}
\end{figure}

A simple fully-connected neural network structure, represented graphically in Fig.~\ref{fig:fcnn}, was trained on the regression task\footnote{The open source packages {\tt tensorflow} and {\tt tflearn} were used to to implement all neural networks and are available from {\tt https://www.tensorflow.org} and {\tt http://tflearn.org}, respectively.}. The network was initialised by setting the biases to zero and the weights to a truncated normal distribution centred at zero with a width of 0.02. A {\tt{tanh}} activation function was applied to the nodes in each layer, as well as an L2 regulariser with weight decay set to 0.001. Dropout \cite{Hinton:2012aa,Ba:2013aa,Baldi:2014AA} was also applied to each layer. While many variations of the network structure were investigated, a systematic hyperparameter tuning was not undertaken due to computational limitations. In general, it was found that fewer hidden units and layers than in the illustrated network led to less optimal minima  of the loss function, while a greater number did not appreciably change the outcome. Dropouts in the range 0.3--0.6 were required to eliminate over-fitting. A range of regularisation prescriptions and hyperparameters, as well as a range of activations including {\tt tanh}, {\tt reLU} \cite{Hahnloser:2000AA,Hahnloser:2001AA,Glorot:2011AA}, and {\tt sigmoid} were studied.
The Adam optimiser~\cite{DBLP:journals/corr/KingmaB14} reached the minimum loss with less training than stochastic gradient descent (SGD), and a loss function based on least absolute deviations (L1) rather than least square errors (L2), performed better.

\begin{figure}
		\includegraphics[width=0.45\textwidth]{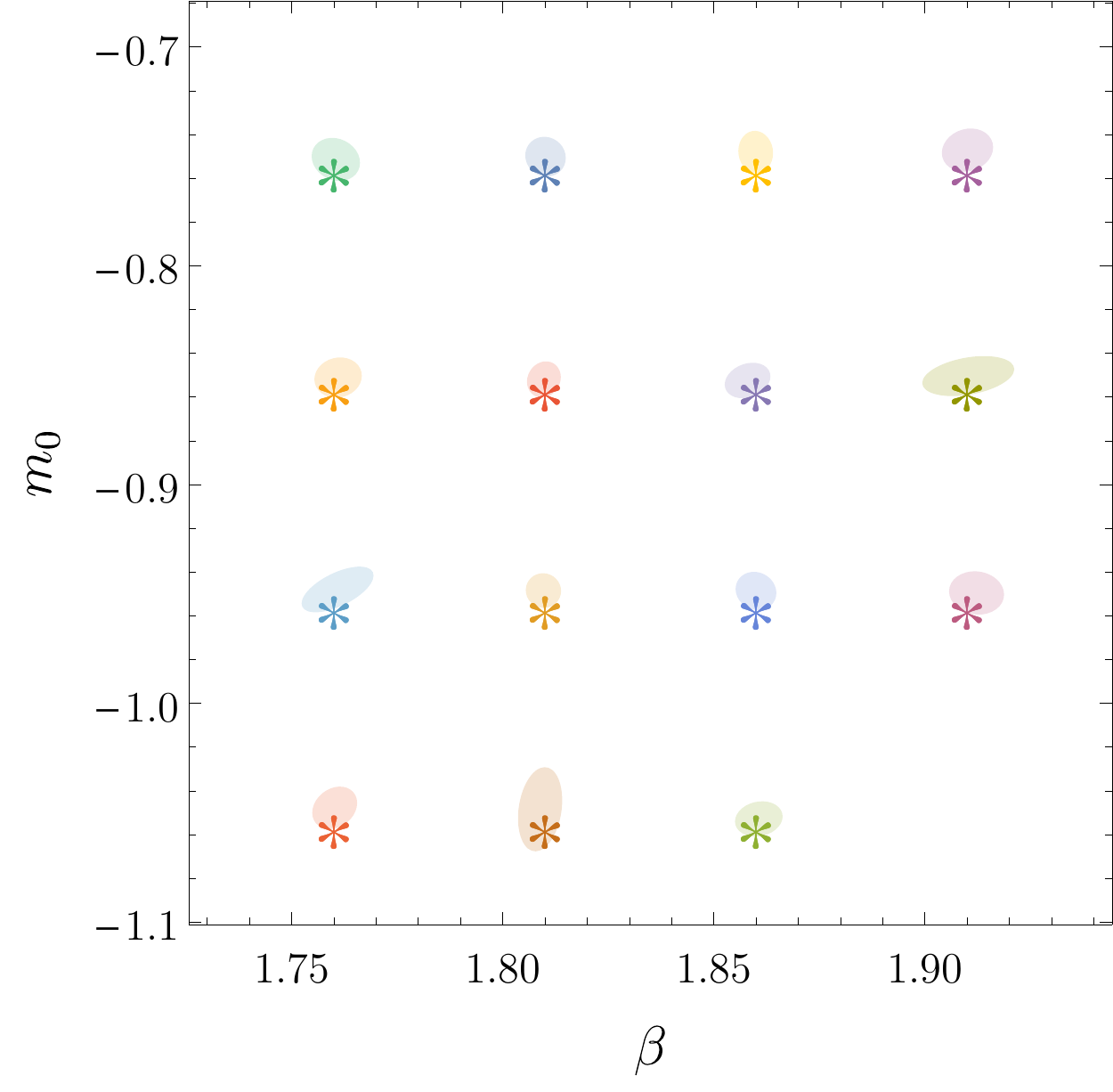}
		\hspace{10mm}
		\includegraphics[width=0.45\textwidth]{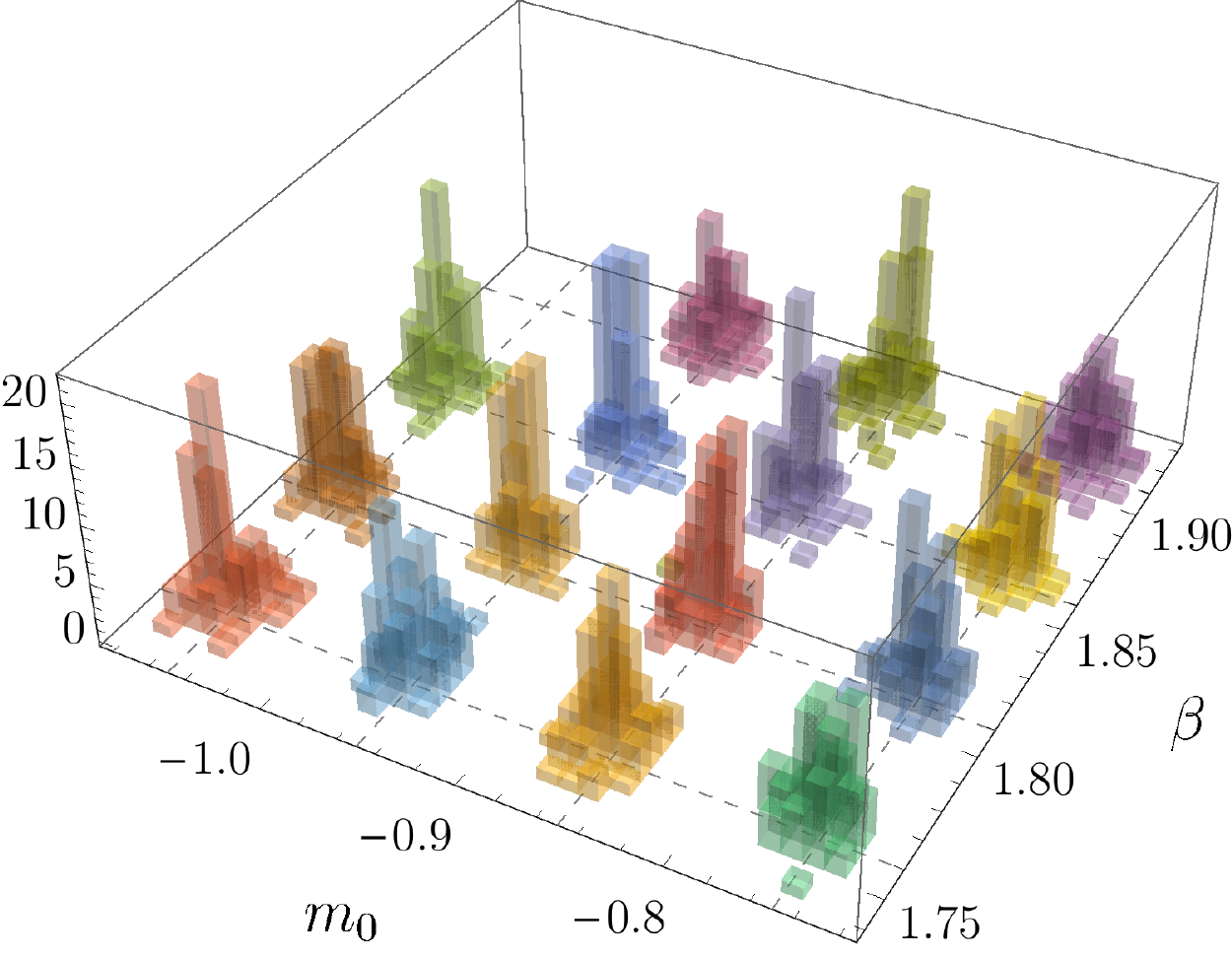}
\caption{\label{fig:valheat}
	Predictions of $\beta$ and $m_0$ on validation ensembles at the same parameter values as the training ensembles. The stars in the left panel denote the parameters used to generate the ensembles, while the ellipses  show the one-standard deviation confidence interval of the predictions for the validation ensembles. The same validation data are shown as histograms in the right figure, with the intersections of the grid lines indicating the parameters used for ensemble generation.}
\end{figure}

The predictions of the best-performing network for the held-out validation data are shown in Fig.~\ref{fig:valheat}. 
While these results appear to signal the success of this approach, the generalisation ability of the network, i.e., its ability to interpolate in parameter space, is poor. In particular:
\begin{itemize}
	\item New ensembles, even those in the 10 ensembles of Set F, generated from separate HMC streams but with the same \{$\beta$,$m_0$\} as one of the training ensembles, were predicted to sit at the average $\beta$ and $m_0$ values of all ensembles included in training. This indicates that the network did not succeed in learning the gauge-invariance properties of lattice QCD gauge fields, nor in parametrising the parameter space of the grid of ensembles;
	
	\item Configurations from the continuation of the HMC streams used to generate the training and validation configurations were also predicted to have different parameters. Specifically, the next configurations in the HMC streams were predicted to have the correct $m_0$ and $\beta$ values, but these predictions drifted towards the average over all training ensembles within a few steps. This indicates that the network is identifying some quantity with a longer autocorrelation time than the physics quantities studied in Sec.~\ref{sec:ensembles}, i.e., that the configurations separated in MC time such that they are independent by the measure of various physics observables, are not independent by the alternative measure found by the network.
\end{itemize}

The majority of these features are unsurprising; information content suggests that with $\mathcal{O}(10^3)$ samples containing  $\mathcal{O}(10^6)$ real numbers each, it is not feasible to stochastically learn symmetries such as the gauge invariance of the data, and that generalisation will be challenging. This could be remedied by using far larger ensembles of gauge field configurations for training, if that were computationally feasible. 

\begin{figure}
	\centering
	\includegraphics[width=0.45\textwidth]{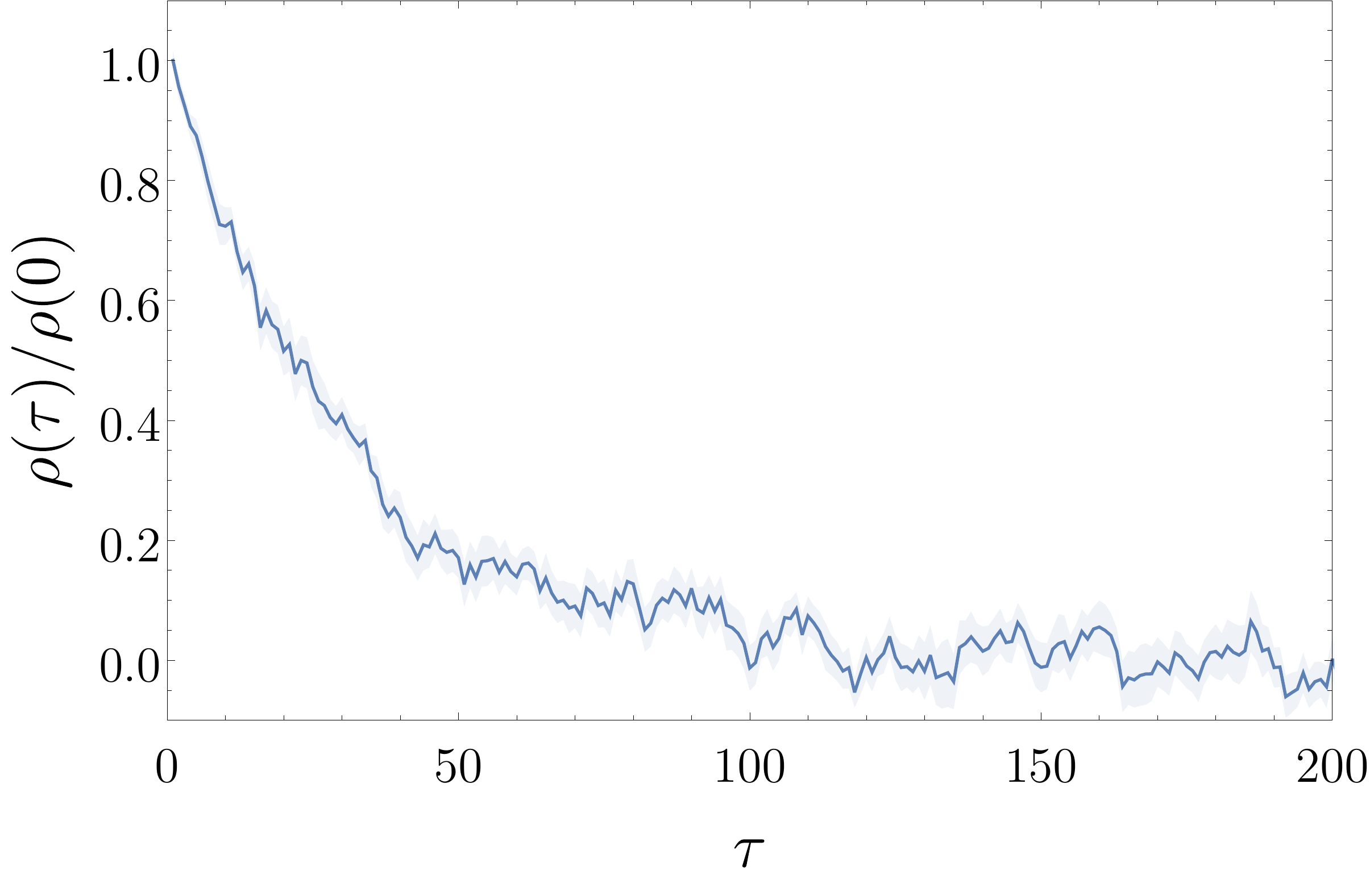}
	\includegraphics[width=0.45\textwidth]{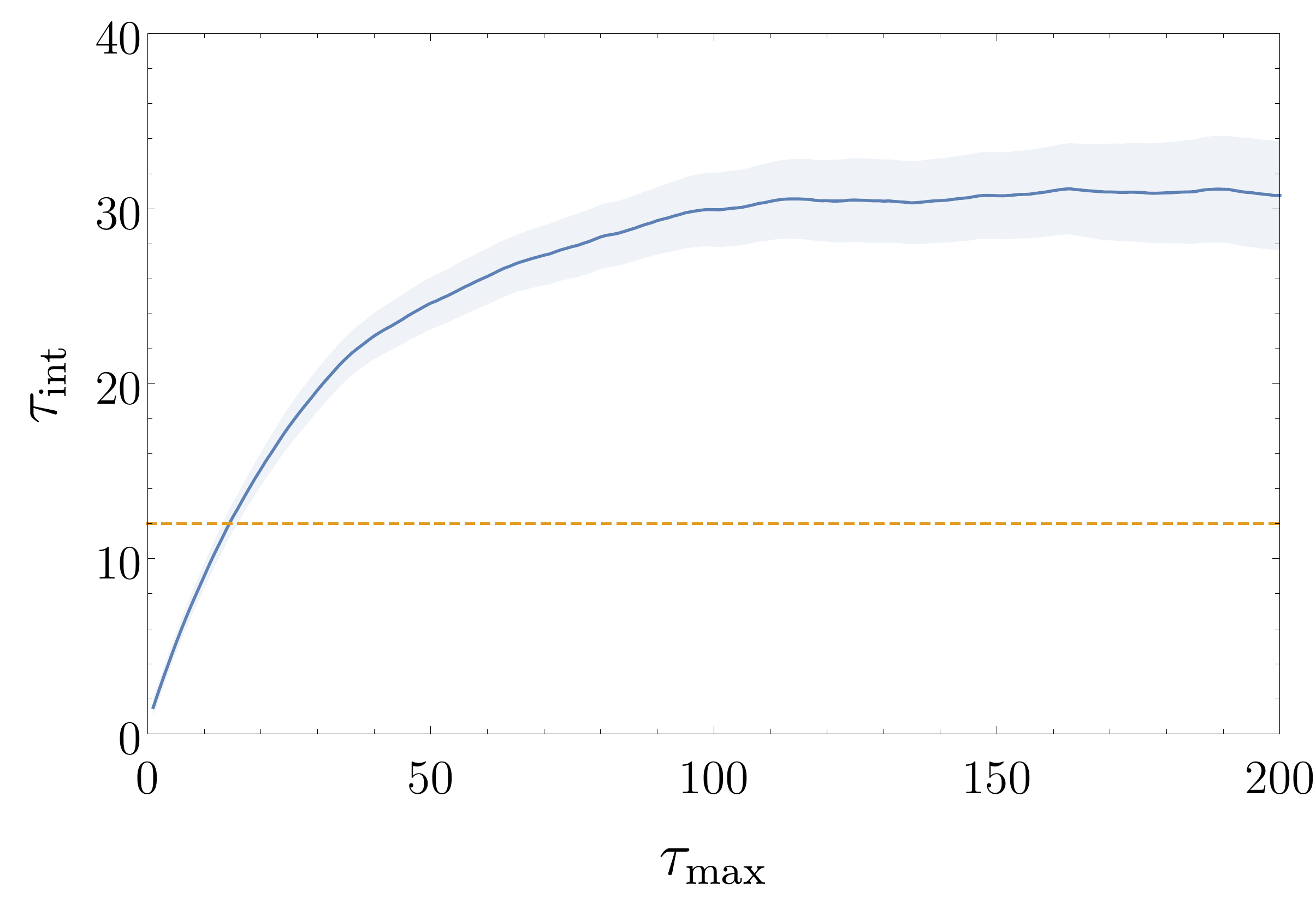} 
	\caption{\label{fig:corrtimeML}
		Autocorrelation function in Monte-Carlo time (left, defined in Eq.~\eqref{eq:MLcf}) and autocorrelation time (right, defined in Eq.~\eqref{eq:tau}) of the feature distinguishing two streams at the same set of parameters, trained on sequences of gauge field configurations. The autocorrelation function was generated by averaging over many different results (trained using all different pairs of the 10 streams, $F_{1,\ldots,10}$, at the same parameters), and was found to be robust under changes of the network structure used to generate it. The dashed horizontal line on the right figure shows the maximum autocorrelation time of various physics observables (see Fig.~\ref{fig:corrtimepirho}). }
\end{figure}

The ability of the network to distinguish different streams generated at the same values of $\beta$ and $m_0$ is interesting. In the limit of infinite stream lengths, no calculated quantity, corresponding to a physical observable or otherwise, can achieve this distinction. Such distinguishability indicates that the streams are not completely sampling the gauge field configuration space and is tied to the existence of a feature, identified by the network, that has a longer autocorrelation than those of the physics observables studied in Sec.~\ref{sec:ensembles}. 
An autocorrelation time of the neural network feature was obtained from the output of classification networks  trained on each of the pairs of streams in Set F, generated at the same set of action parameters. Rather than training this network to identify the $\{\beta,m_0\}$ of a given gauge field as for the regression network described previously, the classifier was trained to produce a classification: $\{1,0\}$ for configurations from one stream, and $\{0,1\}$ for those from a second. The network structure used was identical to that shown in Fig.~\ref{fig:fcnn}, with a {\tt softmax}~\cite{Bishop:2006aa} activation function used for the final layer to provide a normalised probability interpretation for the output: an output $\{a,1-a\}$ for a given configuration indicates that that sample can be identified with the first stream with a probability $a$. A categorical cross-entropy~\cite{36,doi:10.1162/neco.1989.1.3.412} loss function was used for this training. For each pair of streams,
600 trajectories from each stream were used to train an instance of the network. The output of that instance  for the trajectories sequentially following the training data defines an autocorrelation function: 
\begin{equation}\label{eq:MLcf}
\rho(\tau) = 2\left[P_\alpha\left(c^\alpha(\tau)\right)+P_\beta\left(c^\beta(\tau)\right)\right]-1,
\end{equation}
where $P^{\{\alpha,\beta\}}(c)$ denotes the probability for configuration $c$ to be in stream $\alpha$ or $\beta$ (i.e., the first and second element of the network output for that configuration). The configuration is labelled $c^{\{\alpha,\beta\}}(\tau)$ to denote a trajectory from stream $\alpha$ or $\beta$, $\tau$ steps in Monte-Carlo time after the end of the sequence used as training data. The autocorrelation function, and an autocorrelation time determined from this function by Eq.~\eqref{eq:tau}, are shown in Fig.~\ref{fig:corrtimeML}. Comparing to Fig.~\ref{fig:corrtimepirho}, it is clear that the autocorrelation time of the feature used by the network to distinguish streams is approximately three times longer  than the longest autocorrelation time of the physics observables that were calculated in Section \ref{sec:lqcd}.

It is natural to speculate that the strong autocorrelation observed in the neural network output is based on some local features of the data, rather than features encoding the physics of interest\footnote{This is supported by the observation that features with similar autocorrelation times were identified using network structures that respect gauge-invariance, but retain full spatial information.}.
Further investigation did not find evidence for this interpretation; neither Moran's I \cite{10.2307/2332142} nor Geary's C \cite{10.2307/2986645} tests supported the existence of correlated spatial regions in the derivatives of the loss function with respect to inputs. There is also no correlation of these derivatives with known spatially-varying physical quantities such as topological charge density and action density. While the long--correlation-time feature could not be identified in this study, it provides an interesting topic for further study.
In particular, it will be informative to investigate how this scale changes with parameter range, particularly in regions of parameter space where topological charge freezing becomes a difficult problem for simulations.

\subsection{Custom symmetry enforcing network structure}
\label{sec:MLloops}

As described in the previous section, experiments with simple fully-connected neural networks were not successful at parametric regression of lattice QCD gauge fields for the training data sets used in this study. 
This is not unexpected; learning the symmetries of gauge field configurations stochastically is certain to be a challenging task. Symmetries of lattice QCD, however, act to reduce the effective degrees of freedom of the problem, and can be incorporated into the structure and training of neural networks in several ways.  
First, the stochastic learning of symmetries can be accelerated through data augmentation (i.e., randomly performing a gauge transformation and/or translation/lattice rotation on a configuration). This is analogous to typical uses of data augmentation~\cite{NIPS1996_1250} in, for example, image recognition \cite{Baird:1995aa,Wang:2017AA}, to introduce symmetries such as rotational symmetry\footnote{The incorporation of symmetries into various neural network structures has been studied in Refs.~\cite{NIPS1991_536,2014arXiv1411.7817K,Scholkopf:1996aa,NIPS1996_1253}.}. In practice, this was found to be untenable for the case studied here as a result of the large number of symmetries that must be learned, their complex nature, and the requirement that they be strictly observed.
Secondly, custom network layers can be designed (or equivalently, data can be pre-processed) to only allow gauge invariant and lattice-symmetry invariant outputs of the network. This approach is found to be successful.

To incorporate the symmetries of lattice QCD gauge fields into neural network structures, several custom networks were designed, featuring an initial pre-processing layer that forms only quantities that respect the invariances of the problem, followed by fully-connected layers operating on these quantities. The possible gauge and translation-invariant degrees of freedom that are allowed by the first layer are specified by hand; in principle this choice could be part of the learning process, although na\"ive implementations are prohibitively expensive.
Wilson loops of all shapes and sizes, along with their correlated products, suitably averaged over spacetime, provide a natural choice of gauge-invariant, translation-invariant quantities that can be formed from a gauge field configuration. The number of such loops is exponentially large in the spacetime volume and it is computationally intractable to allow all to be generated, so a suitable subset must be chosen. As used in the PCA analysis in Section~\ref{sec:PCA}, one such subset is the set of square planar loops of sizes up to $L/2 \times L/2$, as well as $1\times n$ rectangular loops for $n$ up to $L$, averaged over all possible planar orientations and space-time locations. 
Another natural choice is the set of all correlated products of two Wilson loops, similarly averaged:
\begin{equation}
{\cal W}_{j\times k,l\times m}(R)=\sum_{|r|=R}\sum_{\ell \in {\cal O}(j\times k)} \sum_{\ell^\prime\in {\cal O}(l\times m)} \sum_x{\cal W}_{\ell}(x){\cal W}_{\ell^\prime}(x+r),
\end{equation}
where the sum over $\ell\in {\cal O}(j\times k)$ is over all lattice rotations of loops of size $j\times k$, and these loops are chosen from the same list as the single loops described above. Histograms of these correlated loop products for each ensemble in Grids A, B, and C are shown in Figs.~\ref{fig:looploophistsA}, \ref{fig:looploophistsB}, and \ref{fig:looploophistsC} in Appendix~\ref{app:ensemblevalid}. A third choice of simple, gauge-invariant quantities is the  set of subtracted correlated products of loops, 
\begin{equation}
{\cal W}^{\rm (sub)}_{j\times k,l\times m}(R)=\sum_{|r|=R}\sum_{\ell\in {\cal O}(j\times k)} \sum_{\ell^\prime\in {\cal O}(l\times m)}\left[ \sum_x{\cal W}_{\ell}(x){\cal W}_{\ell^\prime}(x+r) - \sum_x{\cal W}_{\ell}(x)\sum_x{\cal W}_{\ell^\prime}(x)\right].
\end{equation}
Network structures that allow each of these sets---labelled as single loops (SL), unsubtracted products of two loops (CP), and single loops plus the subtracted correlated products of two loops (SLCP)---to be formed in the first layer, are studied.

\begin{figure}
	\centering
	\includegraphics[width=0.8\textwidth]{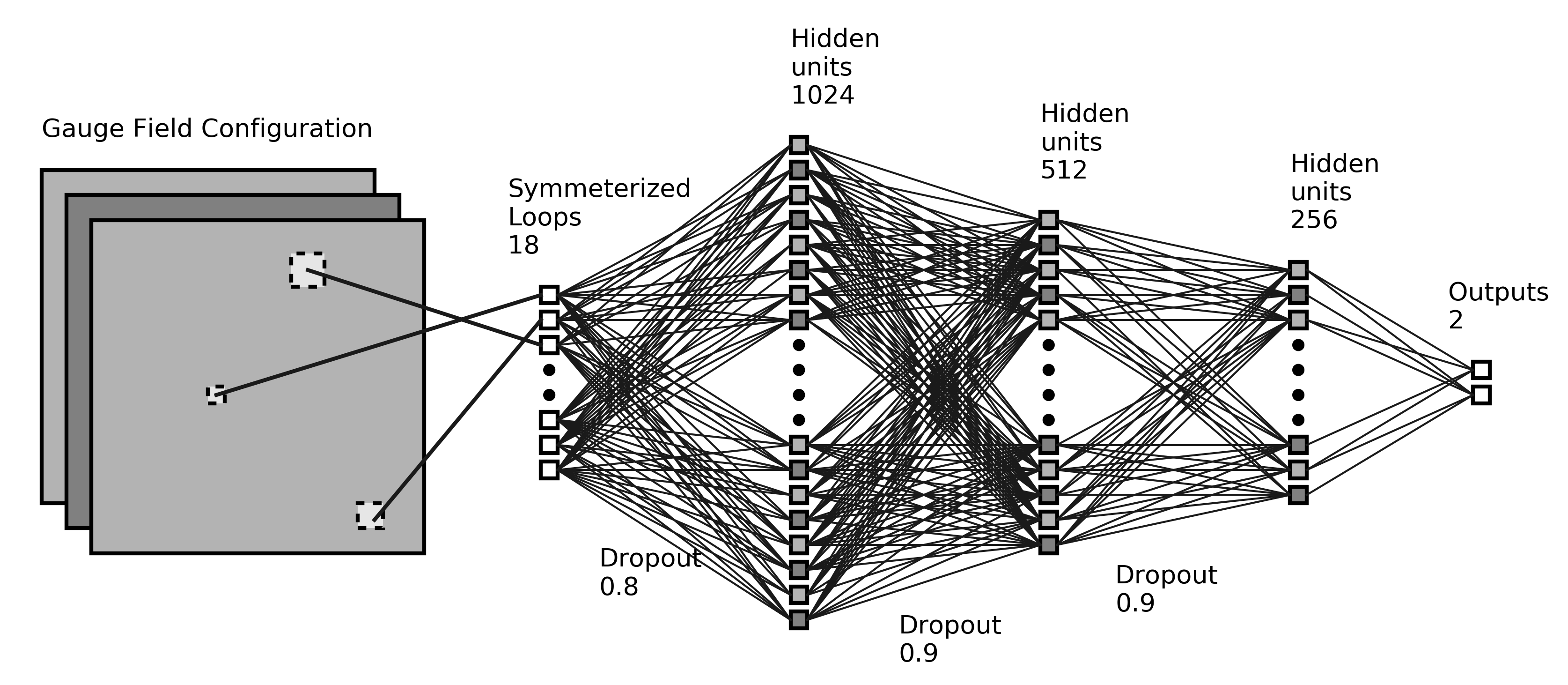} \\ (a) SL\\
	\includegraphics[width=0.8\textwidth]{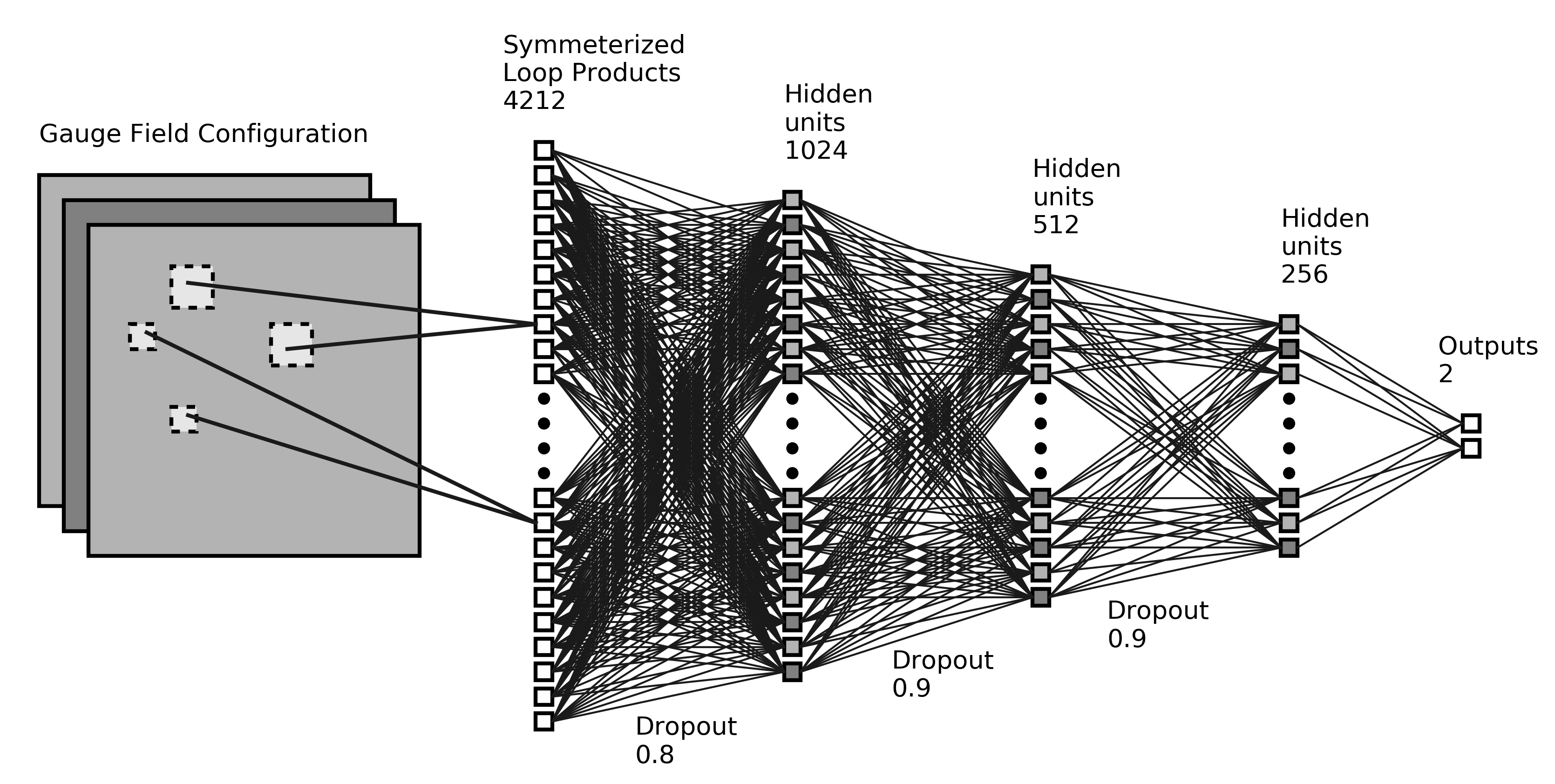} \\ (b) CP\\
	\includegraphics[width=0.8\textwidth]{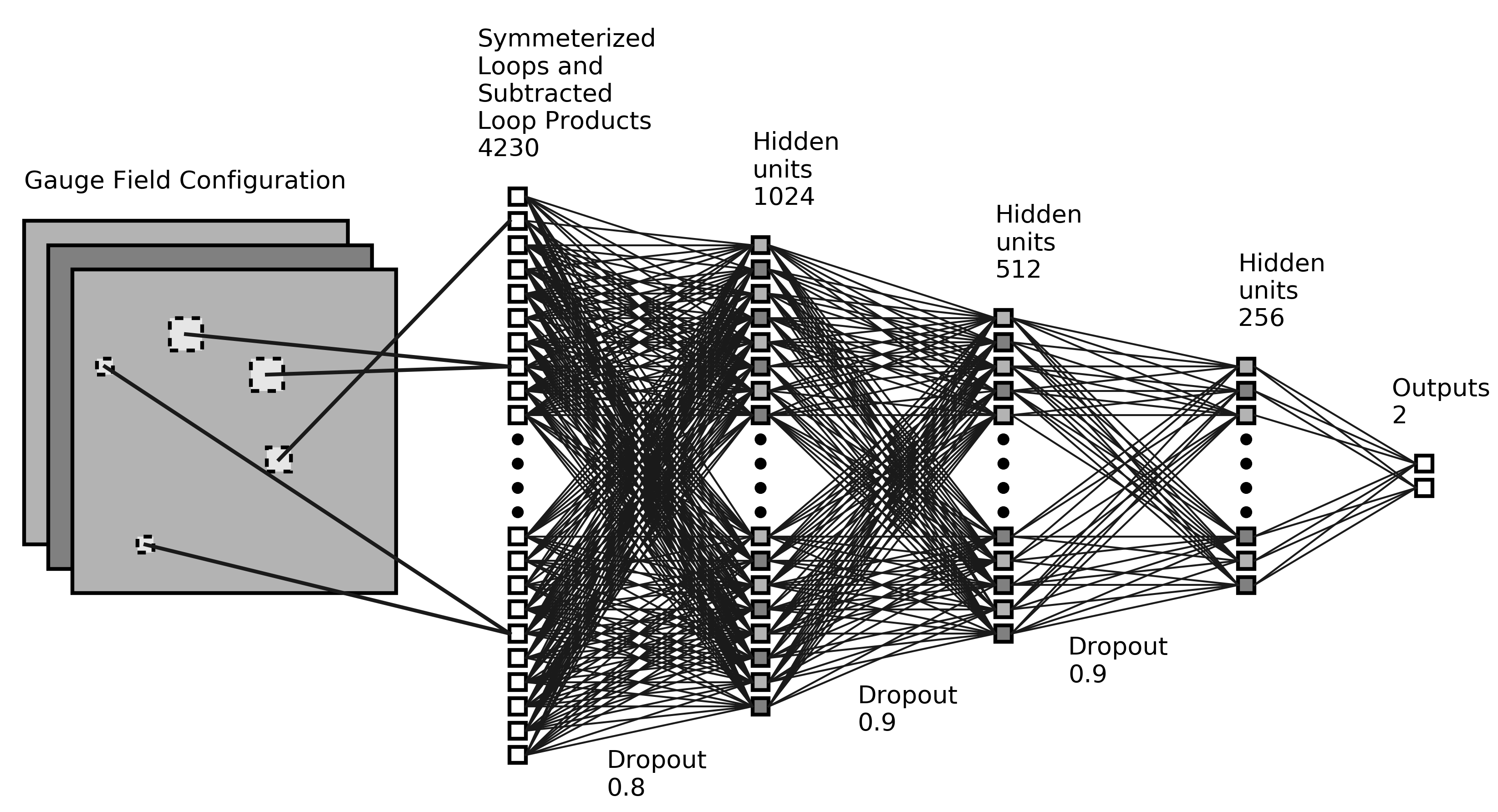} \\ (c) SLCP
	\caption{\label{fig:netstruct}Diagrams of the neural network structure used. In the first layer, SL, CP, or SLCP structures are formed, e.g., in the CP case, products of the 18 different types of loops separated by lattice distance $R < 13$ (averaged in integer space bins of $R$) are allowed, for a total of $18 \times 18 \times 13 = 4212$ loop products. The first layer is followed by 3 fully connected hidden layers with 1024, 512, and 256 nodes. Each hidden layer uses a {\tt tanh} activation function, with dropouts between layers.}
\end{figure}

The complete network structures used for regression are illustrated in Fig.~\ref{fig:netstruct} for each of the SL, CP, and SLCP cases. 
Each network was trained using 850 independent configurations from each ensemble in a given grid, with a further 100 held out as validation data. As for the fully-connected network described in the previous section, the networks were initialised by setting the biases to zero and the weights to a truncated normal distribution centred at zero with a width of 0.02.
Although no rigorous tuning of the hyperparameters of the networks was undertaken for the various structures, a large number of variations were investigated. In general, networks with fewer hidden units, or fewer layers, than those illustrated in Fig.~\ref{fig:netstruct} were found to produce less optimal solutions, while larger networks did not significantly improve on the results that are presented. As for the fully-connected networks, an L1 distance in the two-dimensional parameter space was used as the loss function, and this was found to perform considerably better than the L2 distance. For a given network structure and loss function, the same minimum loss was achieved using different choices of optimiser, including SGD, Adam \cite{DBLP:journals/corr/KingmaB14}, and Nesterov \cite{Nesterov:1983aa}, with various parameters, although the number of epochs required  to convergence varied.

\begin{figure}[!ht]
	\includegraphics[height=0.47\textwidth]{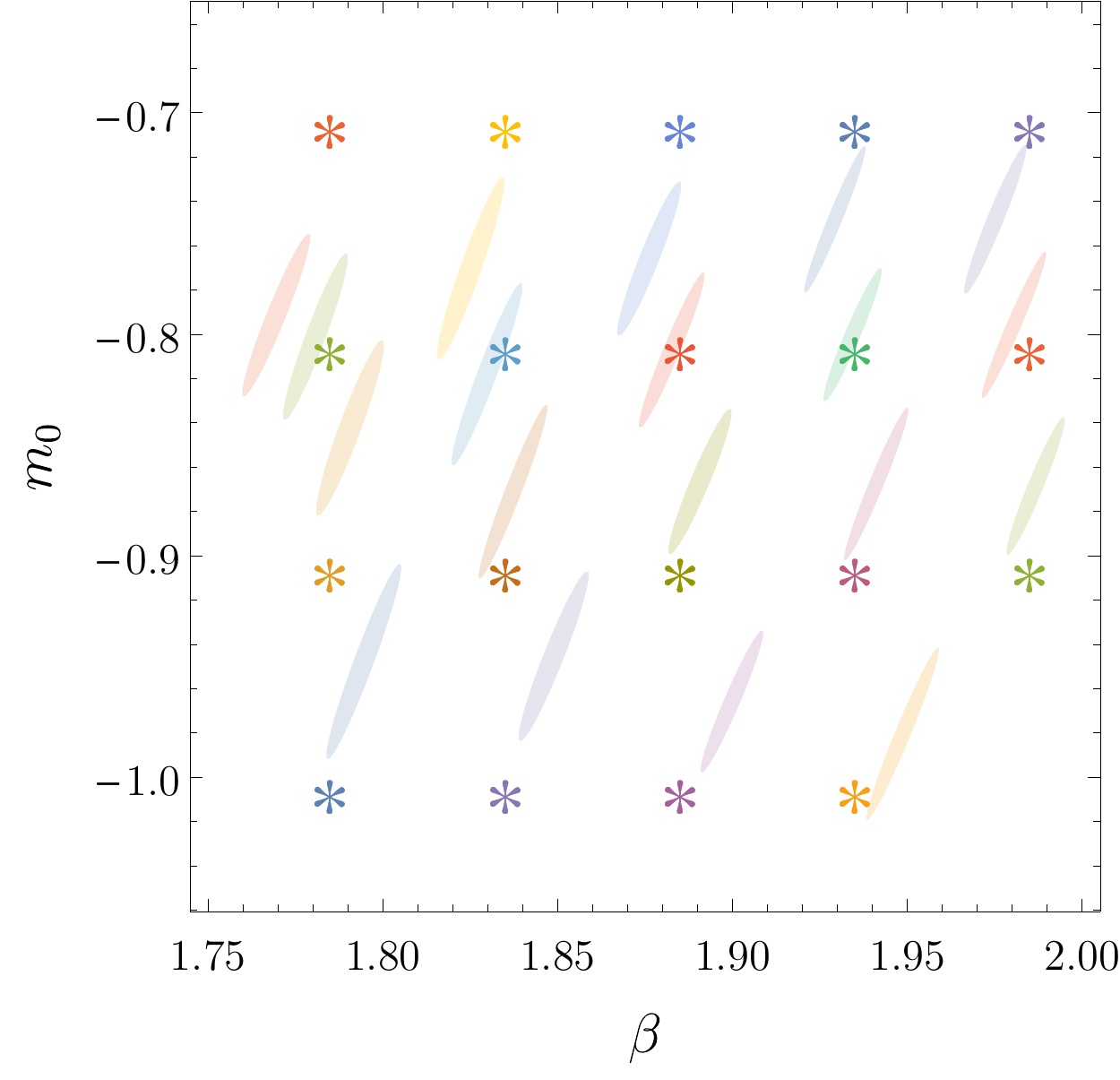}
	\includegraphics[height=0.47\textwidth]{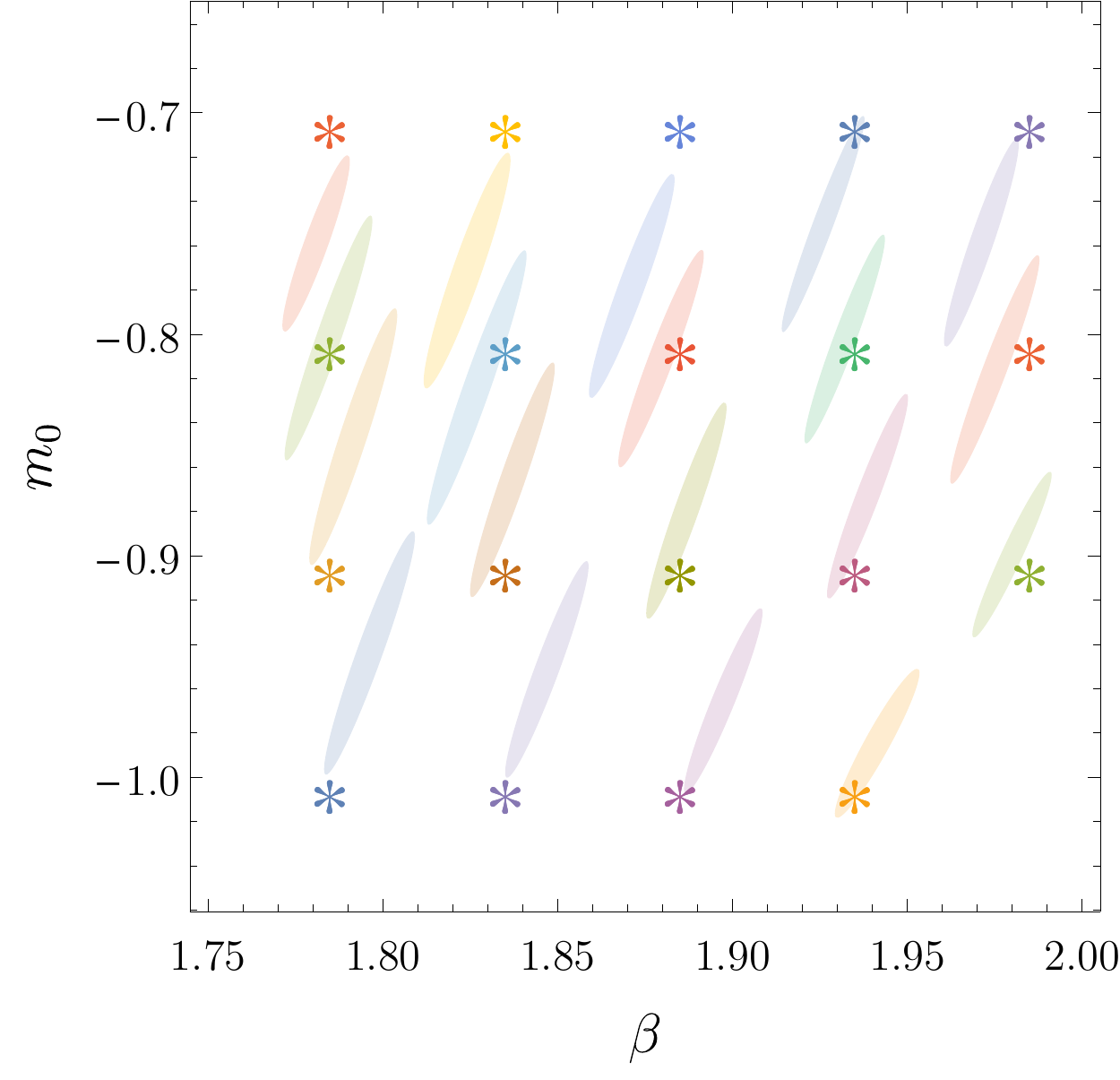}
	\includegraphics[height=0.47\textwidth]{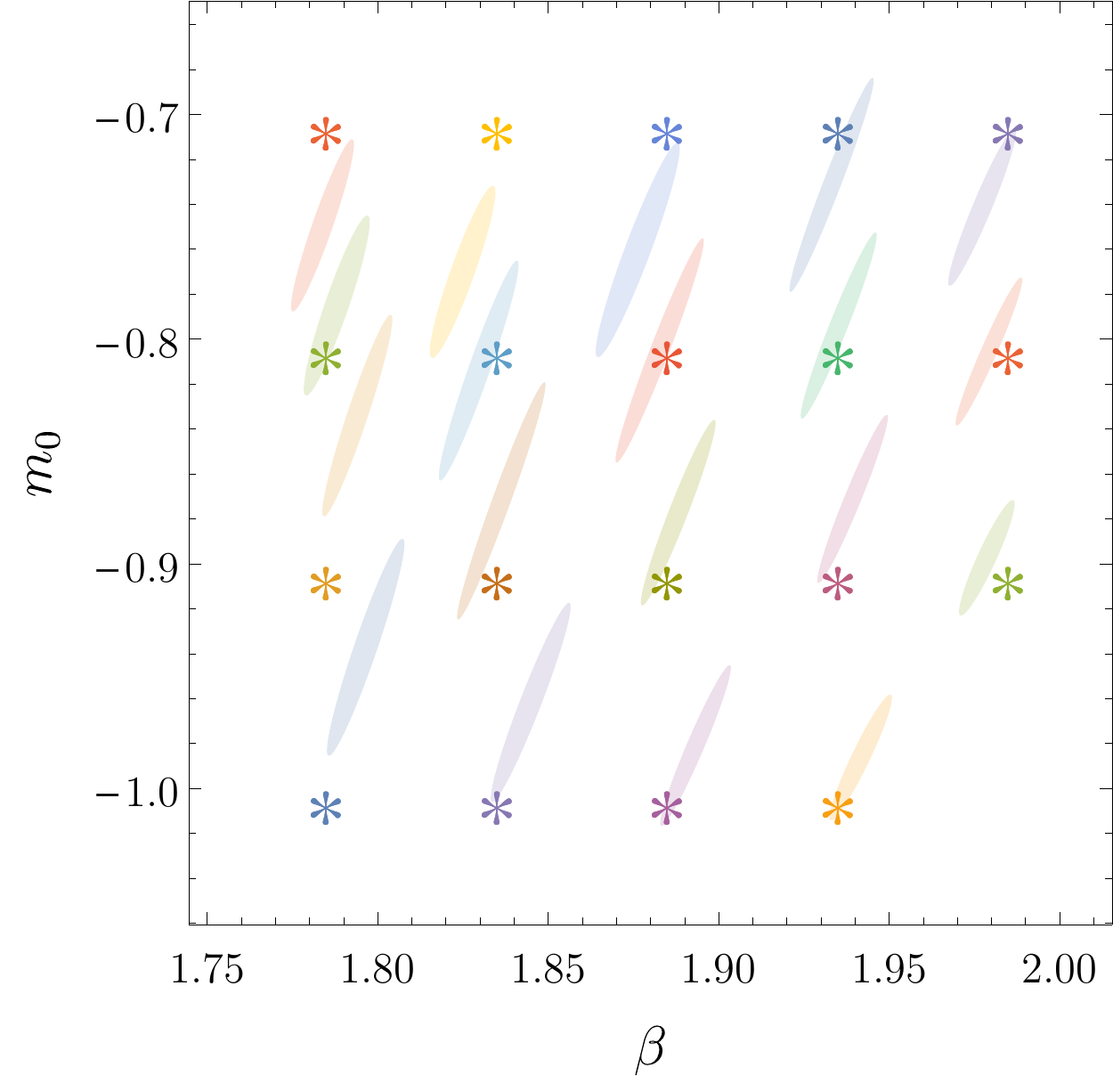}
	\caption{\label{fig:resultsA1}Predictions of $\beta$ and $m_0$ for the validation ensembles in Grid A at the same parameter values of the training ensembles, using SL (left panel), SLCP (right panel) and CP (bottom panel) network structures. The stars show the location of each ensemble in parameter space, while the ellipses show the 1$\sigma$ confidence regions generated from the variation of the predictions for the 100 validation samples from each ensemble. }
\end{figure}

\begin{figure}
	\includegraphics[height=0.4\textwidth]{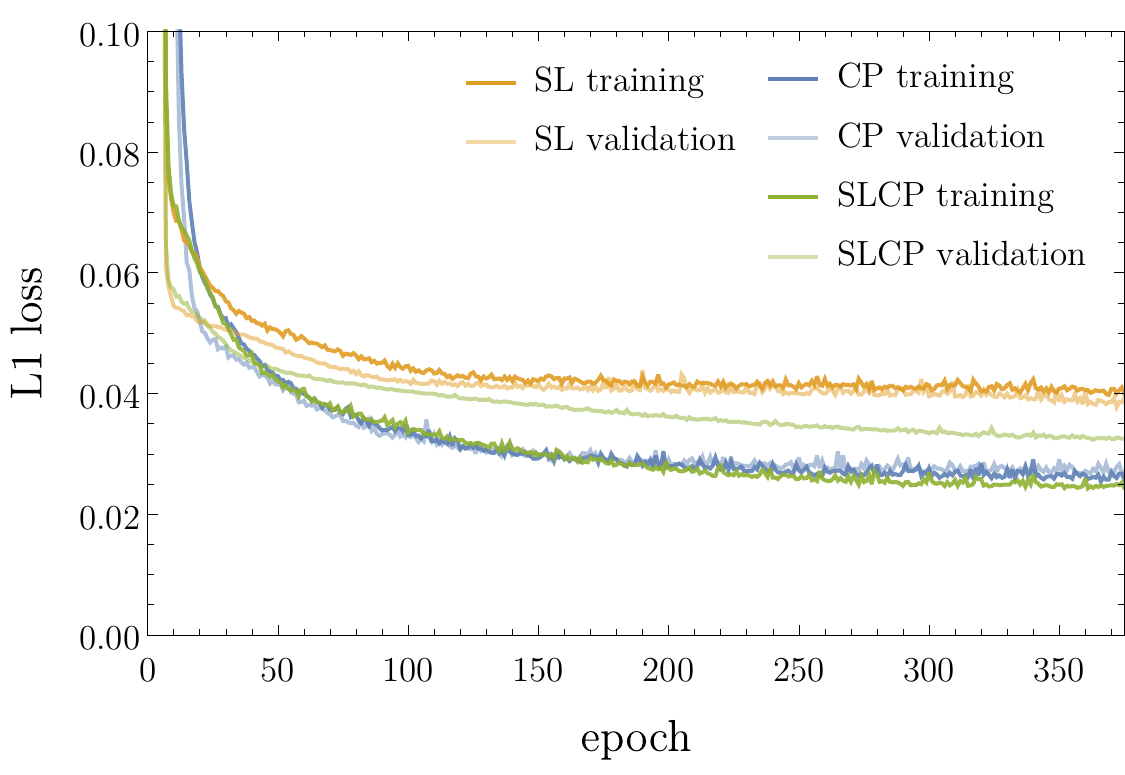}
	\caption{\label{fig:loss}Loss for networks trained on the ensembles in Grid A with SL (orange), CP (blue), and SLCP (green) structures in the first layer, optimised with the Adam optimiser. The dark lines indicate the training loss and the pale lines show loss on the  validation data.}
\end{figure}

The outputs of neural networks allowing each of the SL, SLCP, or CP loop sets to be formed in the first layer, trained on the ensembles in Grid A, are shown in Fig.~\ref{fig:resultsA1}. In each case, the results display accurate regression and clear differentiation between the ensembles, with the shapes of the confidence ellipses of network predictions elongated in the direction of constant $1\times 1$ plaquette, the simplest and most precise gauge-invariant object. {The mild distortion of the regression results towards the centre of the grid is natural, as this will always lead to a smaller loss in the case of misidentifications than any alternative. With additional tuning and larger or denser parameter grids for training, one might expect that this distortion can be removed.} The training and validation losses of each network are shown against training epoch in Fig.~\ref{fig:loss}. The CP network performs slightly better than the SL network, as one may anticipate, given that it allows a larger number of degrees of freedom to be utilised. The SLCP network, while also having more degrees of freedom than the SL network, displays over-fitting: while the training loss is as good as that of the CP network, the validation loss remains higher.
It is likely that tuning the network hyperparameters individually for each network structure would improve these results. For the purpose of the present proof-of-principle study, the CP network is taken as the best example for further study.

\begin{figure}
	\includegraphics[height=0.47\textwidth]{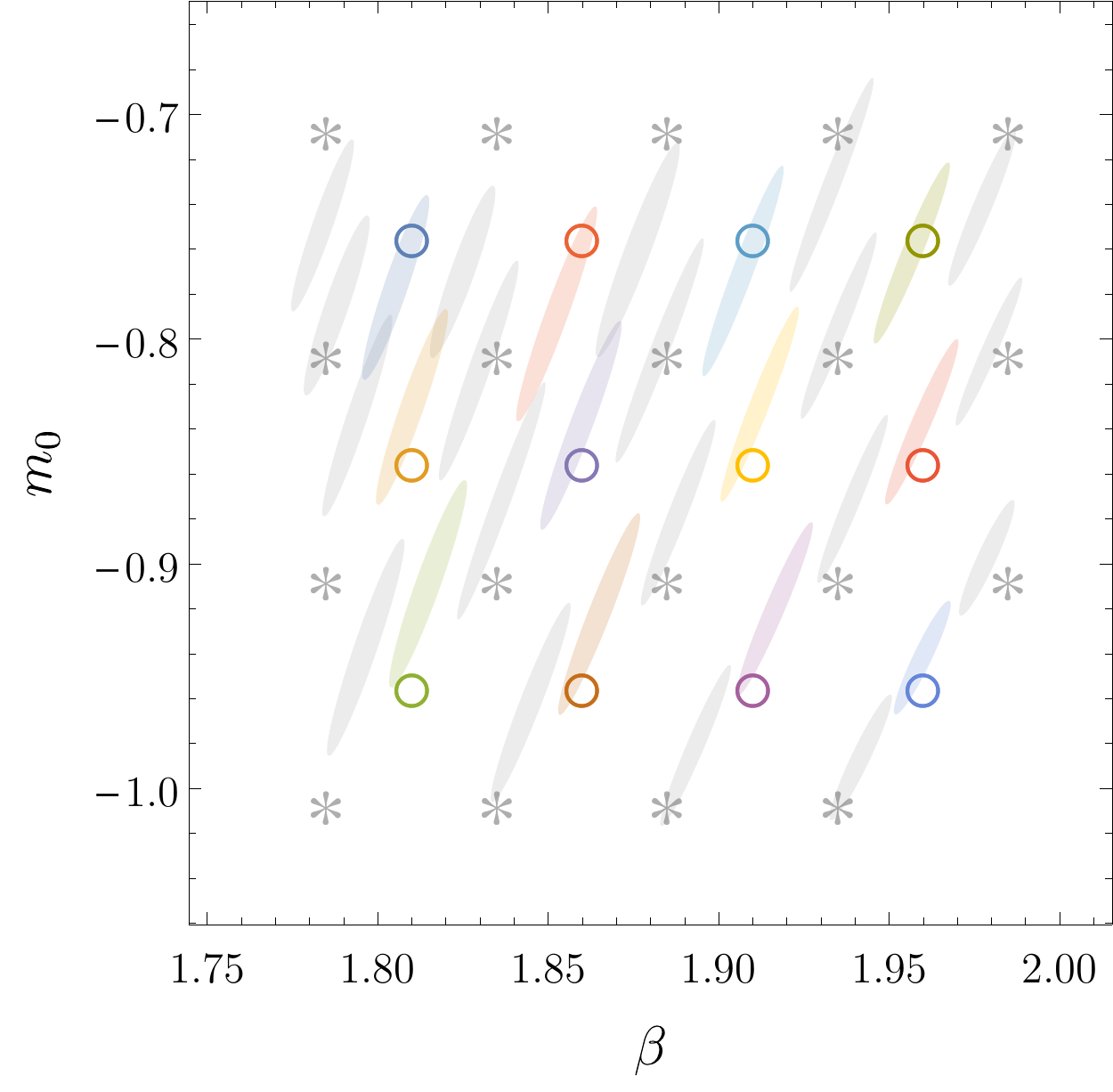}
	\includegraphics[height=0.47\textwidth]{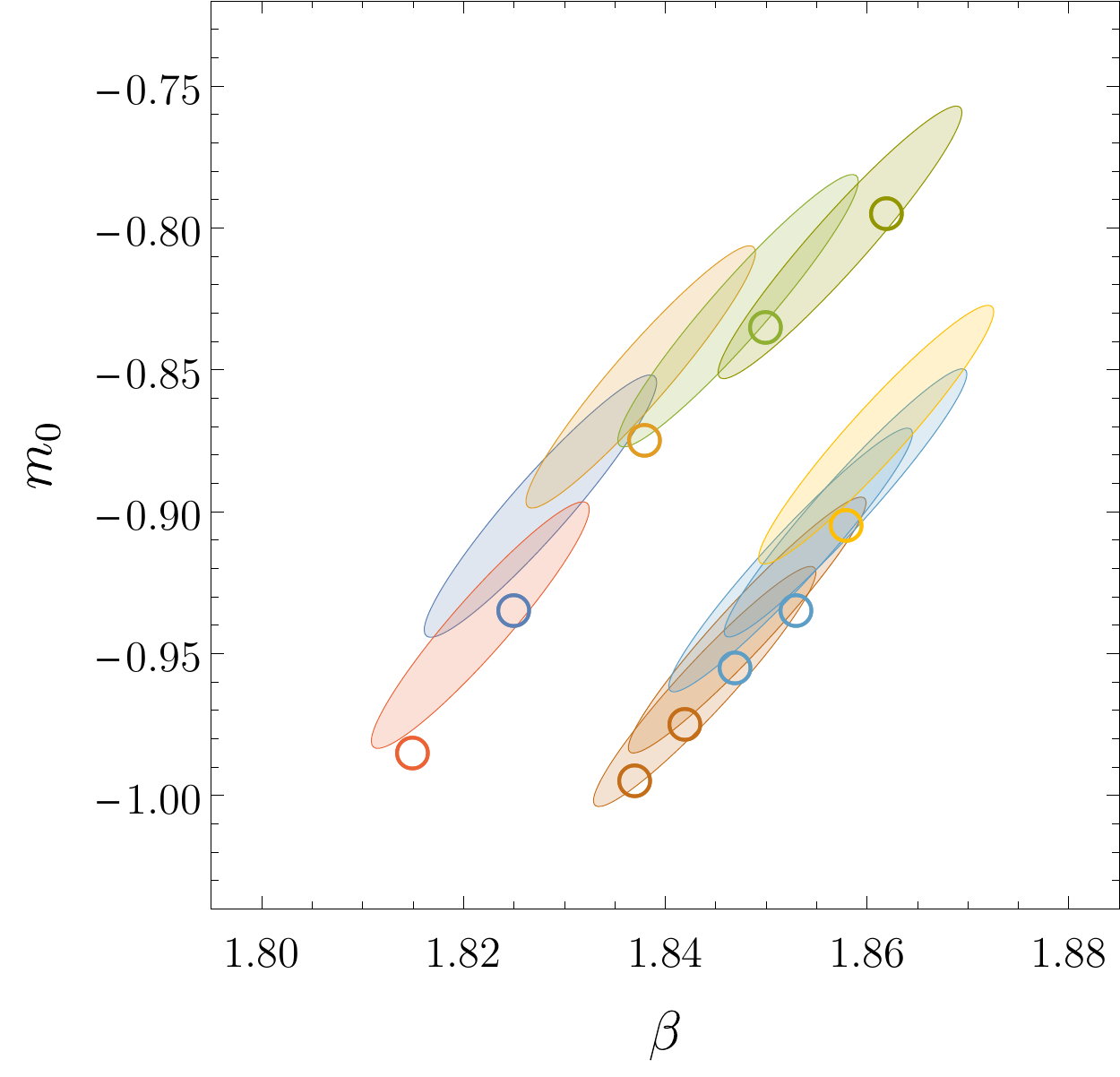}
	\caption{\label{fig:resultsA2}Predictions of $\beta$ and $m_0$ from the CP network trained on Grid A, for the ensembles in Grid B (left panel) and Sets D and E (right panel). The open circles show the the location of each ensemble in parameter space, while the ellipses show the 1$\sigma$ confidence regions generated from the variation of the predictions for the 100 validation samples from each ensemble. The greyed-out stars and ellipses show the validation data and training ensemble locations.}
\end{figure}

Unlike the fully-connected network described in the previous section, the symmetry-respecting networks generalise successfully, both correctly identifying the parameters of other streams generated with the same action as the training data, which are indistinguishable from the validation distributions, and interpolating to intermediate ensembles. This interpolation is illustrated in Fig.~\ref{fig:resultsA2}, which shows the predictions of the CP network on both the evenly-spaced intermediate ensembles of Grid B, and on ensembles in Sets D and E, generated to lie along lines of constant plaquette (isoplaquette lines). While the latter ensembles are essentially indistinguishable along each isoplaquette by various Wilson loops, even using a principal component analysis (see Sec.~\ref{sec:PCA}), the parameter predictions from the trained network are distinguishable, and, most importantly, have the correct relative positions in parameter space. The overlap between the network predictions for the very closely-spaced ensembles from Set E is anticipated; as described in earlier sections, there is a maximum resolution inherent in this regression problem. Nevertheless, the ordering of the central values of the distributions remains robust. This shows accurate regression of dense points in a region of parameter space significantly smaller than the space between adjacent training ensembles, confirming that the network has successfully parametrised the relevant features of lattice QCD gauge fields.

\begin{figure}[t]
	\includegraphics[height=0.47\textwidth]{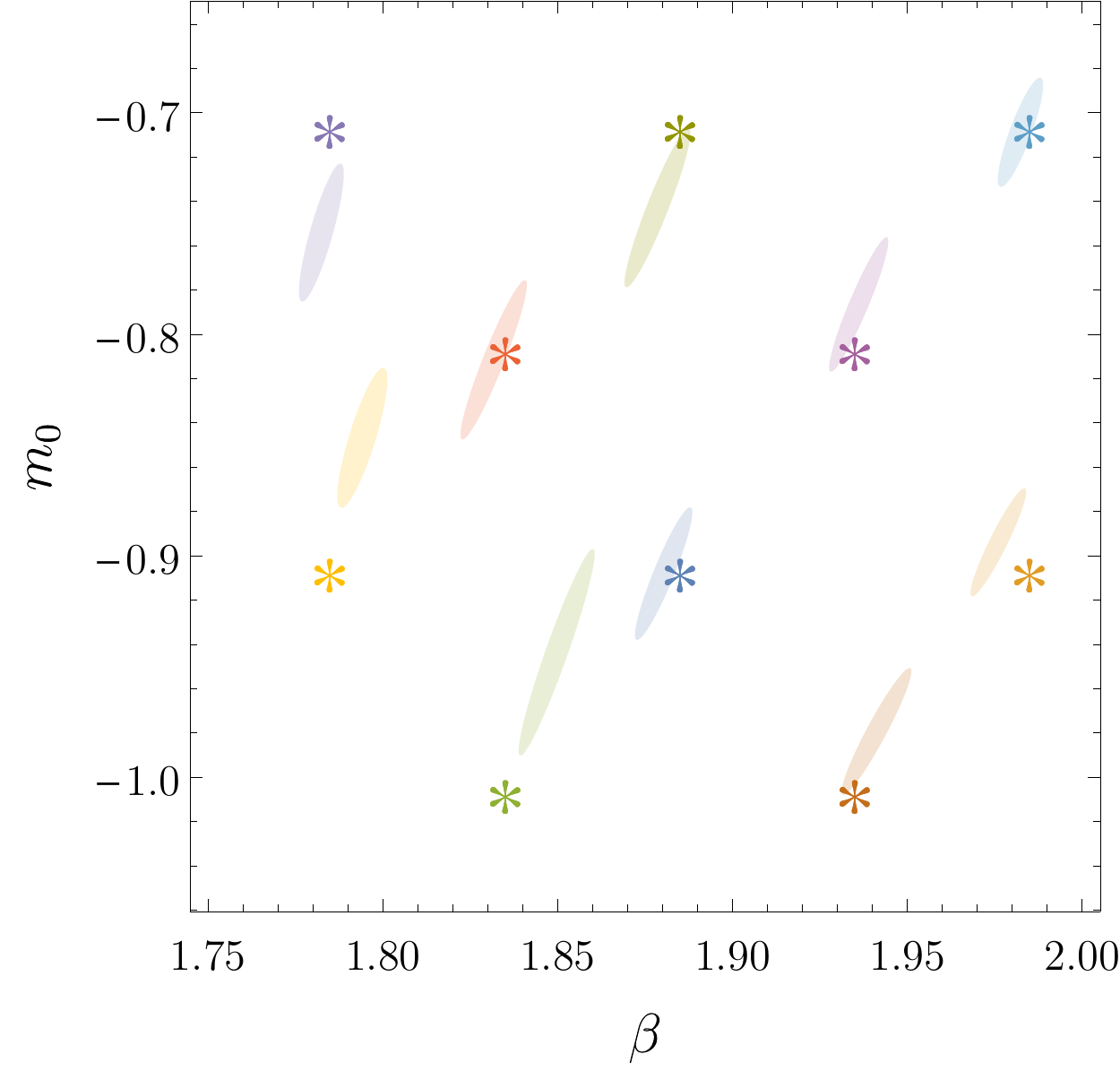}
	\includegraphics[height=0.47\textwidth]{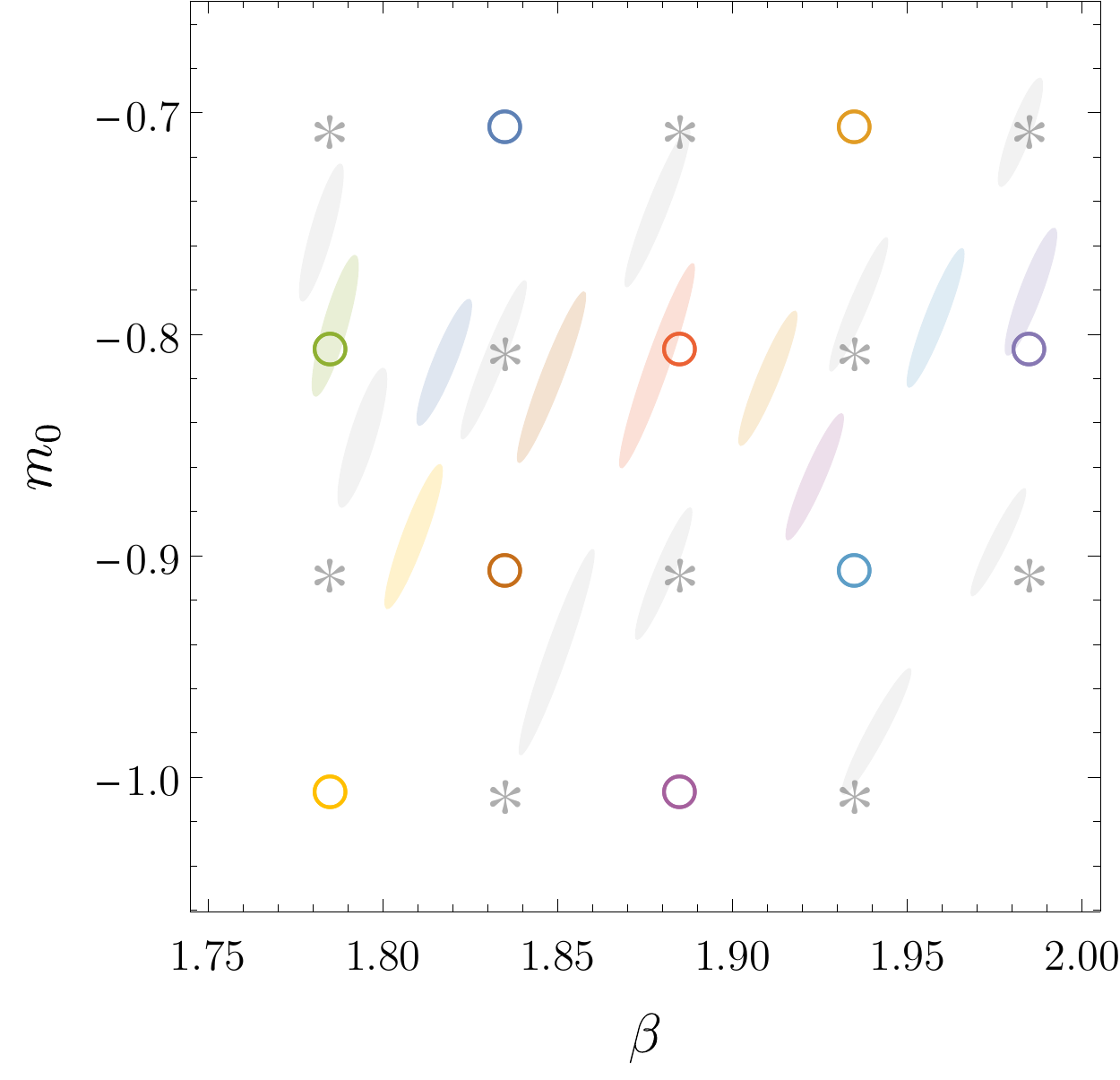}
	\caption{\label{fig:halfdata}Predictions of $\beta$ and $m_0$ from a CP network trained on a subset of the ensembles in Grid A. The stars show the location of each ensemble in parameter space, while the ellipses show the 1$\sigma$ confidence regions generated from the variation of the predictions for the 100 validation samples from each ensemble. In the right panel, the open circles show the location of testing ensembles, that were not included in training, in the parameter space, while the matched-colour ellipses show the 1$\sigma$ confidence regions of the network predictions.}
\end{figure}

The accurate regression achieved with the CP network  relies on having a sufficient density of points in the $\{\beta,m_0\}$ plane in the training data set to enable interpolation. Reducing this density by half, for example, and training the same network structure in the same manner, yields a network instance that generalises poorly to intermediate ensembles. Fig.~\ref{fig:halfdata} shows the results of such a test, using the Grid A ensembles. Despite the poor generalisation performance, both training and validation loss converge to the same values as for the CP network trained on the entirety of Grid A; that is, the training does not indicate over-fitting.

\begin{figure}
	\includegraphics[height=0.47\textwidth]{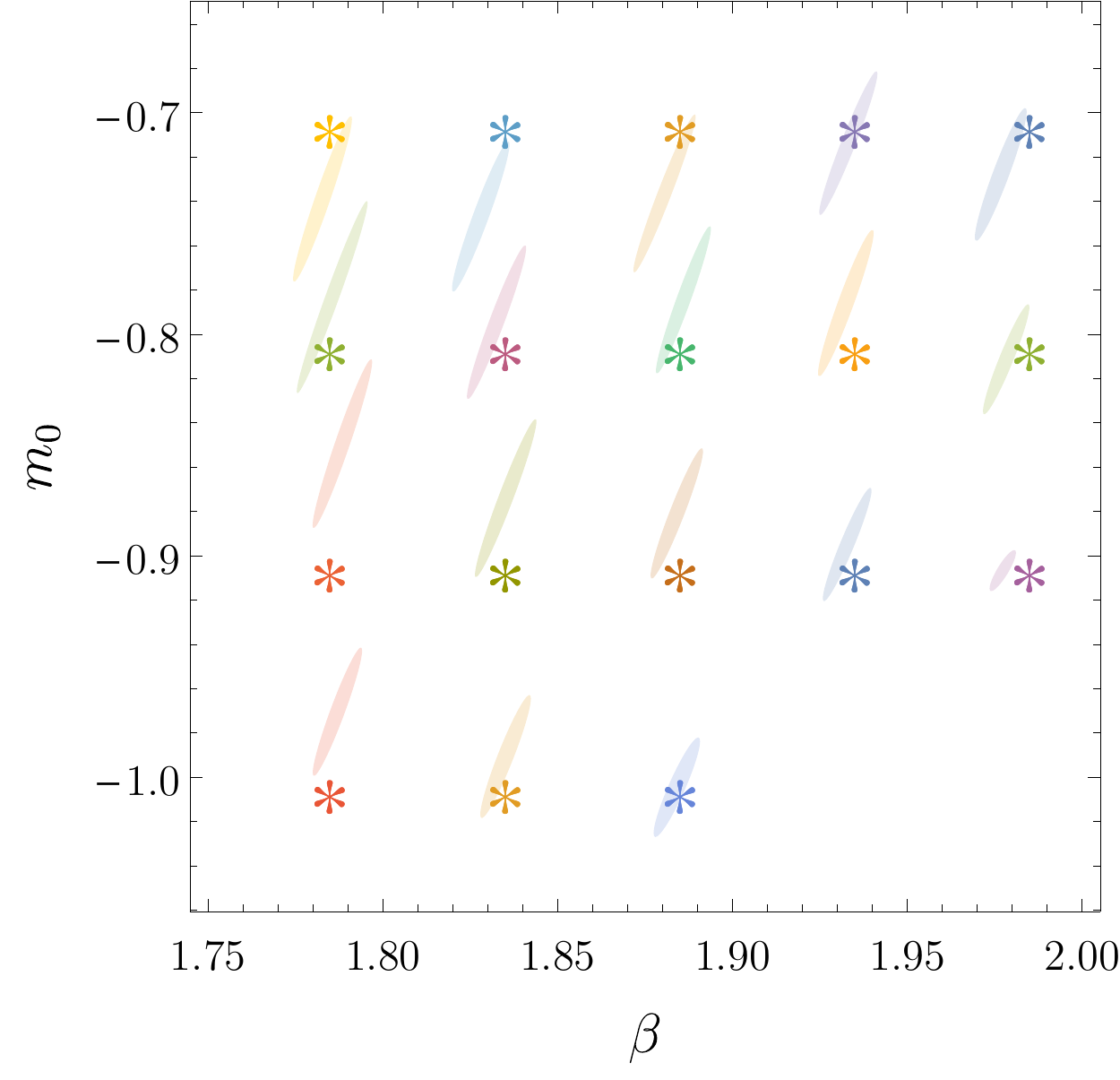}
	\caption{\label{fig:16}Predictions of $\beta$ and $m_0$ for the validation ensembles in Grid C at the same parameter values of the training ensembles, using a CP network structure. The stars show the location of each ensemble in parameter space, while the ellipses show the 1$\sigma$ confidence regions generated from the variation of the predictions for the 100 validation samples from each ensemble.}
\end{figure}

The successful parametric regression of lattice QCD gauge fields presented here must be extended to larger-volume lattices more typical of modern lattice QCD calculations for the method to be applied in practice. As lattice volume increases, Wilson loop distributions become more sharply peaked, and as a result become more distinct, as can be seen by comparing Figs.~\ref{fig:looploophistsA} and \ref{fig:looploophistsC} which display these loops for data sets with spacetime volumes $V=L^3\times T= 12^3\times 36$ and $16^3\times 48$, respectively. It can thus be anticipated that regression performance with the network structures developed here will improve on larger lattice volumes. Fig.~\ref{fig:16} shows the results of a CP network structure trained on Grid C. As expected, the regression performance is better than for the smaller-volume ensembles. Extending these results to even larger volumes, and to $N_c=3$ QCD, is essential.

 \section{Summary}
 \label{sec:summary}
 
 Deep neural networks with custom symmetry-preserving layers provide a solution to the parameter regression problem in lattice QCD, for the proof-of-principle case considered here. 
 Specifically, neural networks regressors trained on grids of ensembles in action parameter space were able to accurately identify the parameters used to generate streams of ensembles, generalising successfully and accurately to ensembles densely spaced and between grid points in the training space. Non-symmetry preserving networks were also studied. While these were unsuccessful at the regression task, they revealed an unknown feature of the lattice ensembles with a longer correlation length than any of the physics observables that were studied.
 
 Extending this work to SU(3) gauge groups and to larger lattice volumes will be essential for the practical application of the methods developed.
 In addition to the symmetries exploited here, a typical length scale, $1/\Lambda_{\rm QCD}\sim 10^{-15}$m, emerges dynamically in LQCD calculations. Consequently, there are potential advantages for a convolutional approach~\cite{Fukushima:1979aa,Fukushima:1980aa,DBLP:journals/corr/GuWKMSSLWW15} at larger lattice volumes. Convolutional layers would again have to be customised, respecting the gauge symmetry of the problem.
 Particular use-cases of LQCD parameter regression may also impose additional constraints. For example, regression for the matching of coarse and fine lattice actions requires the identification of ensembles generated in a coarse space with ensembles describing the same physics, but generated via a coarsening prescription~\cite{Endres:2015yca,Detmold:2016rnh}. The latter ensembles, by renormalisation-group evolution, are described by lattice actions with more parameters than those generated in the coarse space.
 Preliminary investigation suggests that regression under these conditions will require network structures invariant under irrelevant short-distance degrees of freedom, or the marginalisation over  such degrees of freedom in the learning procedure.
 Regression of the larger number of parameters in such actions (and used in the construction of perfect actions~\cite{Bell:1975wtp,Hasenfratz:1993sp,Wiese:1993cb,Hasenfratz:1998gu}), must also be investigated further.
 
 Clearly, having demonstrated the feasibility of neural network approaches to LQCD in the present work, significant further study is warranted. In particular, the use of lattice symmetries to overcome the dramatic inverted data hierarchy of LQCD---the feature that there are typically far fewer samples than degrees of freedom per sample available---opens the door to many novel applications of machine learning in LQCD.

 \begin{acknowledgements} 
 	We are grateful to Kyle Cranmer,  Michael Endres, Brendan Fong, Andrew Pochinsky, and Mike Williams for numerous discussions. The calculations in this project were performed using the
 	Hyak High Performance Computing and Data Ecosystem at the University of Washington, supported, in part, by the U.S. National Science Foundation Major Research Instrumentation Award, Grant Number 0922770, and
 	on clusters at MIT with support from the NEC Corporation Fund. 
 	WD was partly supported by  U.S. Department of Energy Early Career Research Award DE-SC0010495 and grant number DE-SC0011090 and by the SciDAC4 award DE-SC0018121: Computing the Properties of Matter with Leadership Computing Resources. PES and DT were partially supported through contract number DE-AC0506OR23177 under which JSA operates the Thomas Jefferson National Accelerator Facility. DT was supported by the Exascale Computing Project (17-SC-20-SC), a collaborative effort of the U.S. Department of Energy Office of Science and the National Nuclear Security Administration.

 \end{acknowledgements}

\bibliography{ml}


\appendix

\section{Neural Network Glossary}
\label{app:glossary}

{\bf Multi-layer Perceptron} 
A Multi-layer Perceptron is the simplest form of a multi-layer neural network, having a feed-forward network structure, i.e., triggering the activation of each layer of the network successively, without circulating, and consisting of multiple fully-connected layers that use nonlinear activation functions. 

{\bf Loss Function}
A loss, or objective, function, is a measure of the difference between the output of a neural network for a given training sample, and the ground truth. This function defines success for network training. Training procedures, such as stochastic gradient descent, or adaptive learning rate algorithms such as Adam or Nesterov, update the weights and biases of neural networks to minimise the loss.

{\bf Training and validation datasets}
It is typical to hold out some data from training a neural network to form a validation dataset to provide a generalisation test for the network. A larger loss calculated on the validation data than on that used for training is an indication of over-fitting.

{\bf Over-fitting}
The production of a model that is fit to irrelevant features or fluctuations of the training data and therefore fails to generalise reliably.

{\bf Dropout}
Dropout is a regularisation procedure in neural networks whose purpose is to prevent over-fitting. Dropout prevents neutrons from co-adapting by randomly setting a fraction, governed by the dropout hyperparameter, to zero at each training iteration. This results in a model that can be interpreted as randomly sampling from an exponential number of similar networks~\cite{Baldi:2014AA}, and creates more generalisable representations of data.

{\bf Activation}
A neural network layer typically consists of a linear transformation followed by a non-linear transformation at each node, known as the activation function. This non-linearity is what allows neural networks to learn complex decision boundaries. Typical choices of activation functions include {\tt sigmoid}, {\tt tanh}, and {\tt reLU} (defined as $x$ for $x>0$, 0 otherwise).

{\bf Epoch vs. Iteration}
In the training of a neural network, an iteration is one update of the neural net model parameters. Typically, networks are batch-trained, with a hyperparameter governing the batch size of training data considered per update. An epoch is a complete pass through a given training dataset, which may take one (if the batch size is equal to the size of the dataset) or more iterations.

\section{Lattice QCD details}

\subsection{Details of lattice actions, correlation functions, and Wilson loops}
\label{app:lattaction}

The discretised lattice QCD action is expressed in terms of the gauge links between lattice sites, $U_{\mu}(x)$ (which are SU($N_c$) matrices for a theory with $N_c$ colours), and the fermion fields $\psi(x)$, with the Euclidean space-time positions $x\in \Lambda =\{ a(n_1, n_2, n_3, n_4)\big| n_i\in \mathbb{Z}\}$.
The simplest action with the appropriate symmetries for a theory with $N_f$ flavours is given by:
\begin{equation}
S({\beta ,m_{0}}) = \frac{\beta}{N_c} \, \sum_{x\in\Lambda} \sum_{\mu < \nu} \mathrm{Re Tr}  \left[ 1 - P_{\mu\nu}(x) \right] + 
\sum_{f=1}^{N_f} a^4 \sum_{x,y\in\Lambda}
\overline{\psi}_{f}(x)D(m_{0})\psi_{f}(y),
\end{equation}
where $P_{\mu\nu}$ is the plaquette: the shortest, nontrivial, closed loop on the lattice, defined in terms of gauge links as
\begin{equation}
P_{\mu\nu}(x) =U_\mu (x) U_\nu (x+\hat{\mu})U_{-\mu}(x+\hat{\mu}+\hat{\nu})U_{-\nu}(x+\hat{\nu}),
\end{equation}
where $\hat{\mu}$ denotes the vector of length $a$ in the $\mu$ direction.
The Wilson Dirac operator is
\begin{equation}
D(m_{0}) = \left(\frac{4}{a} + m_{0}\right)\mathbb{I} - \frac{1}{a}\sum_{\mu=0}^{3}\left(P_{\mu}^{-}\Omega_{\mu}^{+} + P_{\mu}^{+}\Omega_{\mu}^{-}\right),
\end{equation}
with
\begin{equation}
P_{\mu}^{\pm}  = \frac{1}{2}(1\pm \gamma_{\mu}), \quad \langle x|\Omega_{\mu}^{+}|y\rangle = \delta_{x+\mu,y}U(x,\mu), \,\,\,\,\Omega_{\mu}^{-} = (\Omega_{\mu}^{+})^{\dagger}.
\end{equation}

The action is parameterised by two values: the coupling constant $\beta$ and the bare quark mass $m_0$.

The plaquette can be generalised to Wilson loops of arbitrary shapes and dimensions. Planar Wilson loops $W_{k\times l}(x)$, with indices $k$ and $l$ denoting the dimensions of the loop (with orientation label suppressed), as illustrated in Fig.~\ref{fig:loopfig}, are expressed in terms of gauge links as
\begin{align}\nonumber
W_{k\times l}(x) = &U_\mu(x) U_\mu(x+\hat{\mu})\ldots U_\mu(x+ (k-1)\hat{\mu})\\\nonumber
& \times U_\nu(x+k\hat{\mu})U_\nu(x+k\hat{\mu}+\hat{\nu})\ldots U_\nu(x+k\hat{\mu}+(l-1)\hat{\nu})\\\nonumber
& \times U_{-\mu}(x+k\hat{\mu}+l\hat{\nu})U_{-\mu}(x+(k-1)\hat{\mu}+l\hat{\nu})\ldots U_{-\mu}(x+\hat{\mu}+l\hat{\nu}) \\
& \times U_{-\nu}(x+l\hat{\nu})U_{-\nu}(x+(l-1)\hat{\nu})\ldots U_{-\nu}(x+\hat{\nu})
\end{align}

Two-point correlation functions are defined as the matrix elements corresponding to the creation of some state at a time $0$, and annihilation at some later time $t$. 
For the pion and rho mesons considered in this work, with suitable choices of creation and annihilation operators, the zero-momentum projected correlation functions can be defined as
\begin{equation}
C_{\pi(\rho)}(t)=\sum_{\bf x}\langle 0 | \overline u \gamma_{5(3)} d({\bf x},t)\overline{d}\gamma_{5(3)} u({\bf 0},0)|0\rangle,
\end{equation}
where $u$ and $d$ denote quark creation (and $\overline{u}$ and $\overline{d}$ annihilation) operators. For further details, see Refs.~\cite{Gattringer:2010zz,Rothe:1992nt}.

\subsection{Further details of ensemble properties}
\label{app:ensemblevalid}

\begin{figure}
\includegraphics[width=\columnwidth]{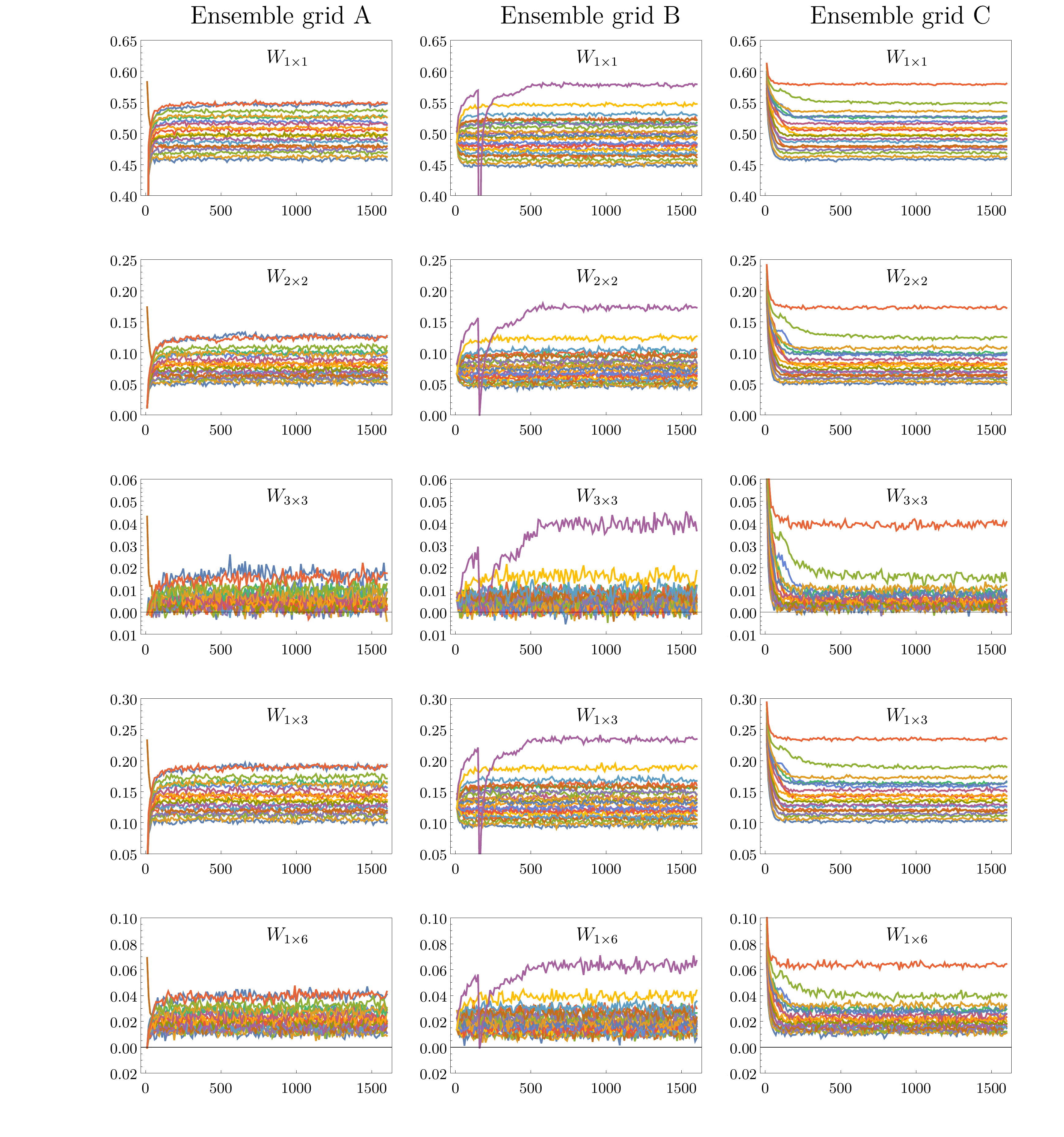}
\caption{\label{historiessummary}
	The various Wilson loops, $W_{m\times n}$, on the first 1600 (of 10000)  trajectories of each of the ensembles in Grid A (left column), Grid B (middle column) and Grid C (right column).
	}
\end{figure}

\begin{figure}
		\includegraphics[width=\textwidth]{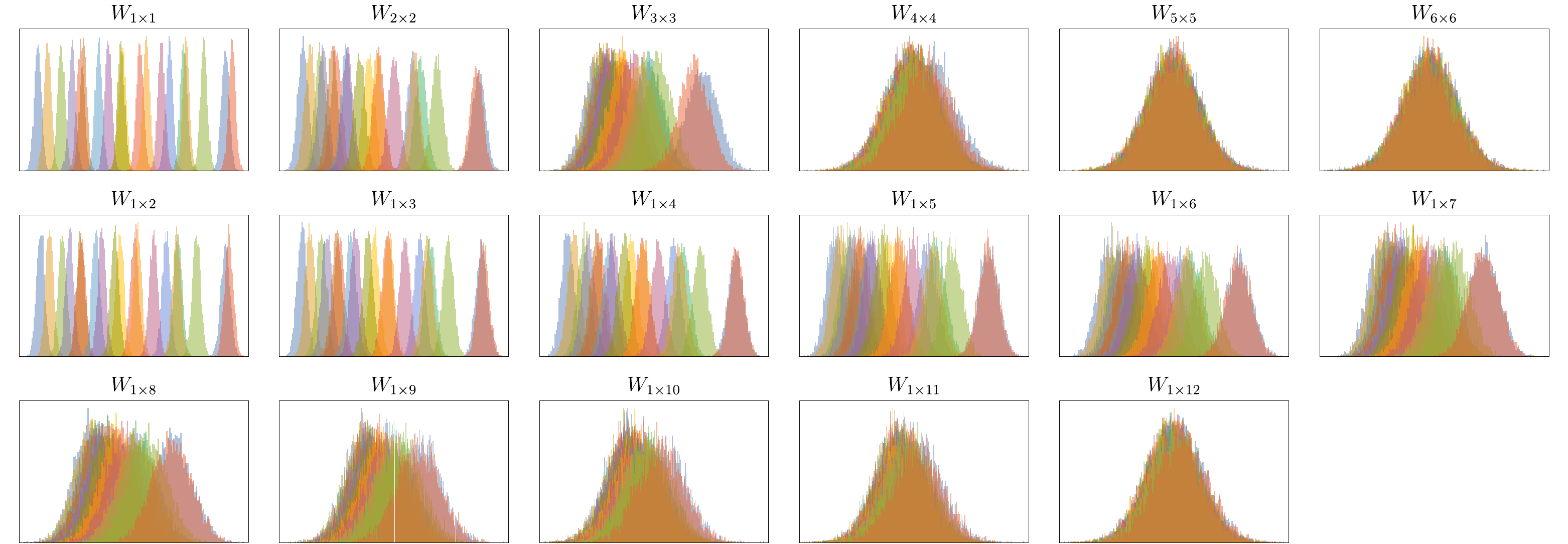}
	\caption{\label{fig:loophistsA}
		The various Wilson loops, ${W}_{m\times n}$, on each of the ensembles in Grid A for all $m,n$ combinations used in this work.
	}
\end{figure}

\begin{figure}
	\includegraphics[width=\columnwidth]{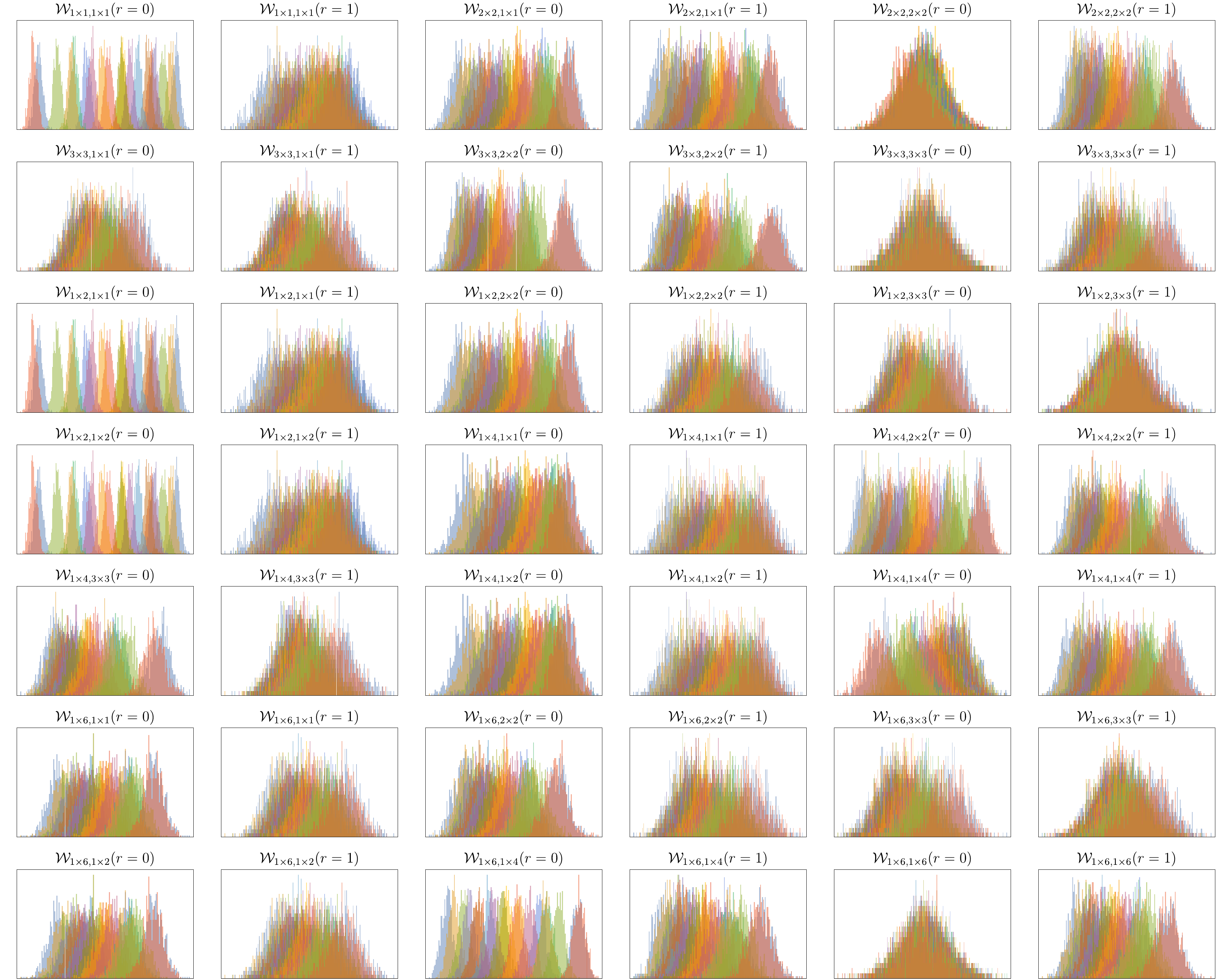}
	\caption{\label{fig:looploophistsA}
		The various Wilson loop correlators, ${\cal W}_{m\times n,p \times q}(r)$, on each of the ensembles in Grid A for a selection of choices of loop shapes and separations $r=0,1$.
	}
\end{figure}

\begin{figure}
	\includegraphics[width=\textwidth]{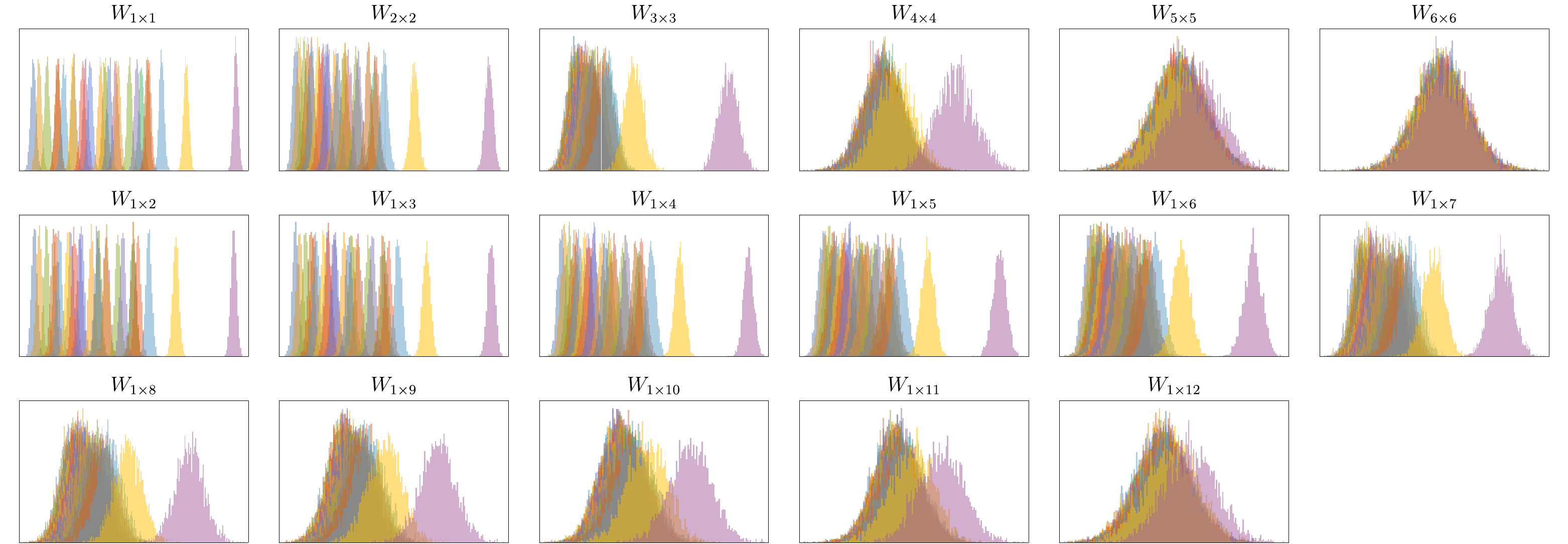}
	\caption{\label{fig:loophistsB}
		The various Wilson loops, ${W}_{m\times n}$, on each of the ensembles in Grid B for all $m,n$ combinations used in this work.
	}
\end{figure}

\begin{figure}
	\includegraphics[width=\columnwidth]{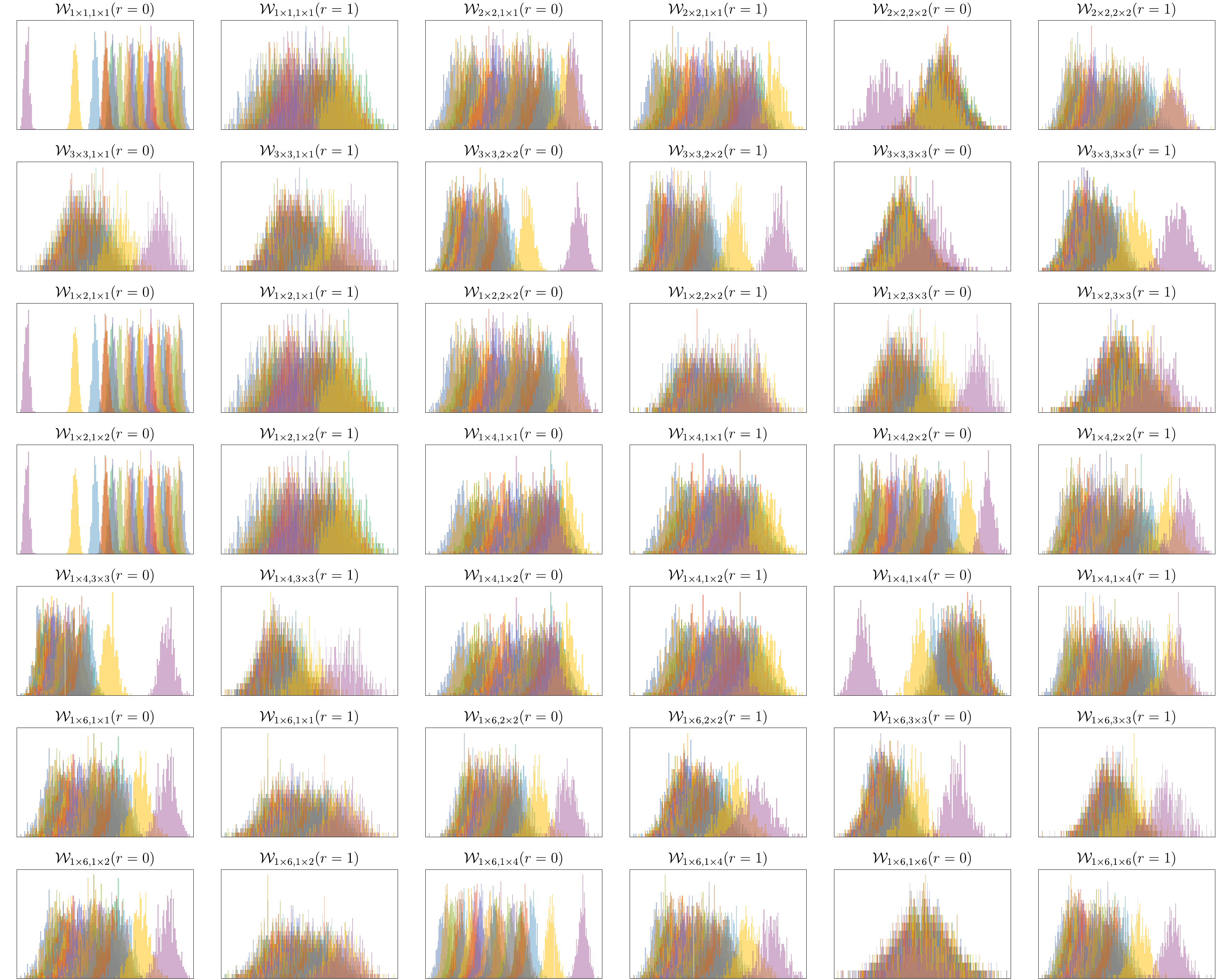}
	\caption{\label{fig:looploophistsB}
		The various Wilson loop correlators, ${\cal W}_{m\times n,p \times q}(r)$, on each of the ensembles in Grid B for a selection of choices of loop shapes and separations $r=0,1$.
	}
\end{figure}

\begin{figure}
	\includegraphics[width=\textwidth]{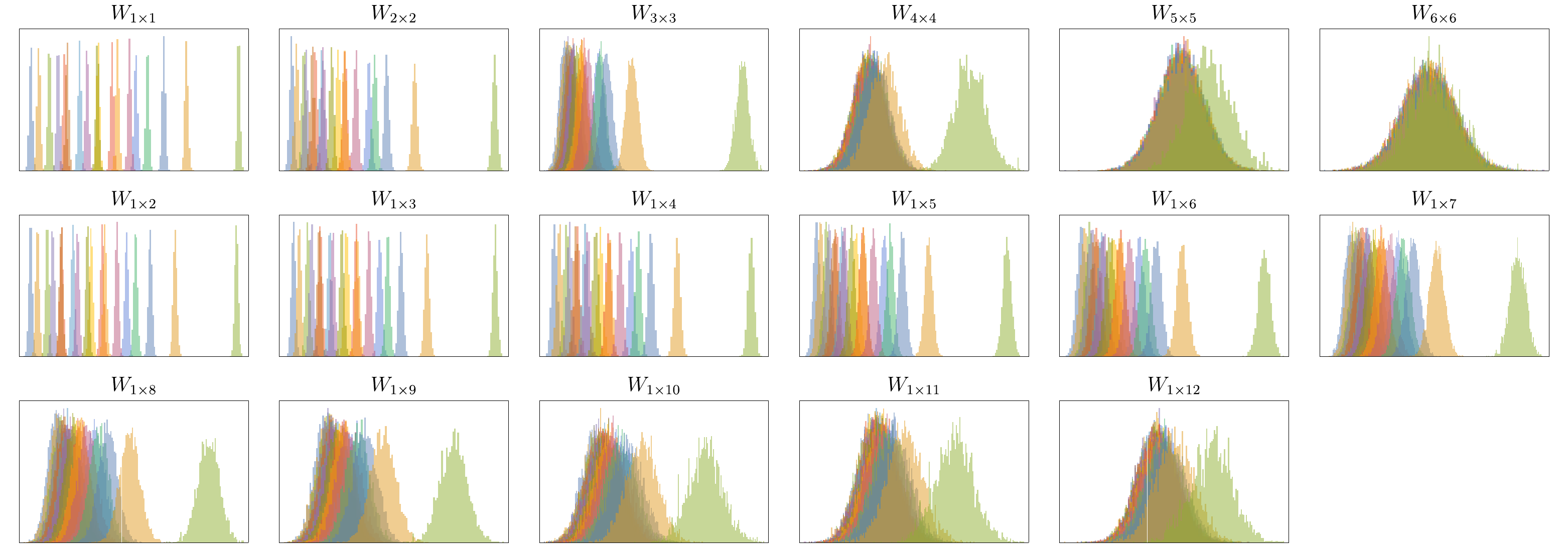}
	\caption{\label{fig:loophistsC}
		The various Wilson loops, ${W}_{m\times n}$, on each of the ensembles in Grid C for all $m,n$ combinations used in this work.
	}
\end{figure}

\begin{figure}
	\includegraphics[width=\columnwidth]{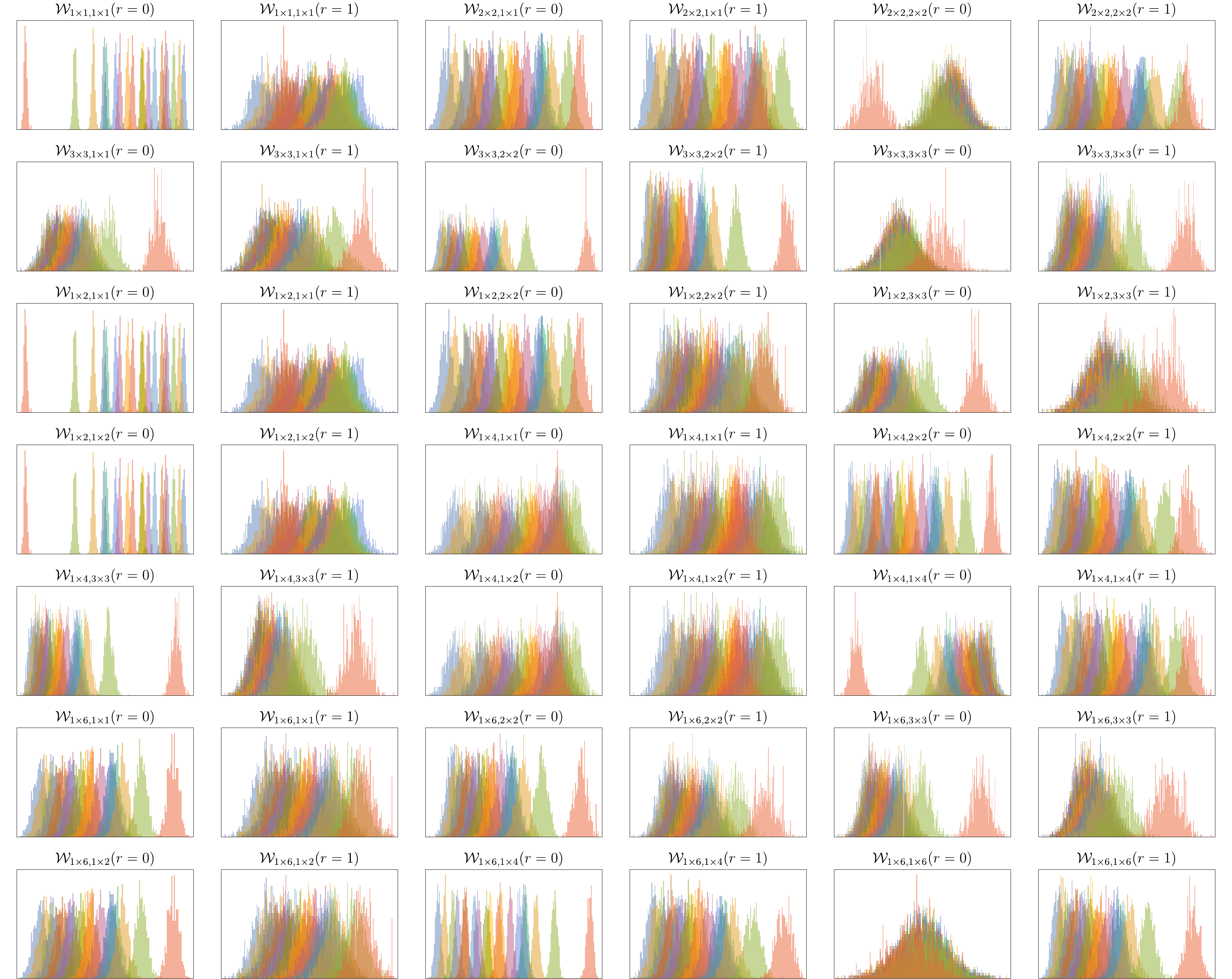}
	\caption{\label{fig:looploophistsC}
		The various Wilson loop correlators, ${\cal W}_{m\times n,p \times q}(r)$, on each of the ensembles in Grid C for a selection of choices of loop shapes and separations $r=0,1$.
	}
\end{figure}

In this appendix, the properties of the various LQCD data sets used in this work are presented. Fig.~\ref{historiessummary} shows the evolution of various Wilson loops with HMC trajectory for the ensembles in Grids A, B, and C, while Figs.~\ref{fig:loophistsA}--\ref{fig:looploophistsC} present histograms of the Wilson loops and correlated products of Wilson loops on each ensemble in these grids.


\end{document}